\documentclass[twocolumn,trackchanges]{aastex701}

\usepackage{mathrsfs}
\usepackage{amsmath}
\usepackage{xcolor}
\usepackage{booktabs}

\begin{document}

\title{Distinct stellar-parameter dependencies of granulation and mesogranulation in red giants}

\author[orcid=0000-0002-4282-605X,sname='Mallick']{Anohita Mallick}
\affiliation{Department of Astronomy and Astrophysics, Tata Institute of Fundamental Research, Mumbai, India}
\email[show]{anohita.mallick@tifr.res.in}

\author[orcid=0009-0008-4832-9936, sname='Verma']{Jharnesh Verma}
\affiliation{Department of Astronomy and Astrophysics, Tata Institute of Fundamental Research, Mumbai, India}
\email{jharneshverma@gmail.com}

\author[orcid=0009-0002-8330-3391, sname='Bhattacharya']{Shatanik Bhattacharya}
\affiliation{Department of Astronomy and Astrophysics, Tata Institute of Fundamental Research,  Mumbai, India}
\email{shatanik.bhattacharya@tifr.res.in}

\author[orcid=0000-0003-2896-1471, sname='Hanasoge']{Shravan M Hanasoge}
\affiliation{Department of Astronomy and Astrophysics, Tata Institute of Fundamental Research,  Mumbai, India}
\affiliation{Center for Astrophysics and Space Science, New York University Abu Dhabi, United Arab Emirates}
\email{hanasoge@tifr.res.in}

\begin{abstract}
Granulation and mesogranulation represent different spatial and temporal scales of convection, but are commonly assumed to follow the same stellar atmospheric scaling relations in red giants. We test this assumption using the largest sample assembled for this purpose to date -- 9,328 Kepler red giants spanning the red giant branch and red clump phase, cross-matched with APOGEE stellar parameters. We model each star’s power spectrum background to obtain the radius-corrected amplitude and characteristic frequency of both components, and derive empirical scaling relations with effective temperature, surface gravity, and metallicity through Bayesian model comparison. Gaussian Process and symbolic regression provide independent validation. We find that granulation and mesogranulation are not self-similar: their surface-gravity dependencies are statistically indistinguishable, consistent with a shared envelope geometry, but their temperature dependencies diverge: mesogranulation's amplitude rises steeply with $T_{\rm eff}$, whereas granulation's shows no statistically justified dependence, and both frequencies respond weakly and similarly. We propose that this pattern is consistent with granulation forming in a thin near-surface layer, where radiation takes over from convection and the Mach number depends only weakly on $T_{\rm eff}$. Mesogranulation originates deeper, where convection carries the full flux and its Mach number and cell size increase with $T_{\rm eff}$. Metallicity raises both amplitudes but has little effect on the frequencies. Our results indicate that the self-similarity assumption underlying two-component background models does not hold, with direct implications for background modeling and its downstream application in Galactic archaeology.
\end{abstract}


\section{Introduction}

Convection is one of the most fundamental physical processes shaping the observable properties of stars. In stars with an outer convective envelope, hot plasma rises to the photosphere, radiates its excess heat, cools, and sinks back into the interior, producing small stochastic brightness fluctuations as convective cells form, evolve, and dissolve across the visible surface \citep{Herschel1801}. This pattern, granulation, imprints itself on time-series photometry as a smooth, non-oscillatory excess of power, the ``convective background" \citep{Harvey1985,Mathur2011}, ubiquitous in stars with convective envelopes and central to the analysis of high-precision photometry from CoRoT \citep{Baglin2006}, $Kepler$ \citep{Borucki2010}, K2 \citep{Howell2014}, and TESS \citep{Ricker2015}.

Red giants are among the primary tracers used in Galactic archaeology, since precise asteroseismic masses, radii, and ages can be obtained for large populations across the Galaxy. Because the oscillation parameter $\nu_{\rm max}$ underlying these determinations is measured from the same power spectrum in which the granulation background must first be modeled and removed, an accurate physical understanding of this background carries direct practical consequences. \citet{Mathur2011} related granulation timescale and power to $\nu_{\rm max}$ and to fundamental stellar properties, demonstrating that the background carries genuine physical information rather than being pure noise. As larger samples became available, it became clear that a single Harvey-like component is insufficient to describe the red giant background; at least two components - ``granulation" and ``mesogranulation," analogous to solar terminology are required \citep{Kallinger2014}.

The physical reality of mesogranulation as a distinct convective mode has been debated in solar physics for decades. Since \citet{November1981} proposed it as an intermediate scale linked to helium ionization at depth, some studies reported detections \citep[e.g.,][]{November1987,Title1989,Muller1992}, while others found no evidence for a distinct scale even in high-resolution data \citep[e.g.,][]{Straus1992,Rieutord2000}. Recent reviews favor the interpretation that no distinct mesogranular scale exists, and that the apparent power instead reflects a continuous distribution of convective scales \citep{Rincon2018,Zhang2025}.

This debate has not been carried into the red giant literature. Since \citet{Kallinger2014}, fitting red giant power spectra with separate granulation and mesogranulation components has become standard practice \citep[e.g.,][]{Corsaro2014,Kallinger2019,Nielsen2021}, typically motivated by fit quality alone. Empirical scaling relations, however, have mostly been derived for the granulation amplitude and characteristic frequency, using large samples from the main sequence to the late red giant branch \citep{Mathur2011, Kallinger2014,Kallinger2016, Corsaro2017}. The mesogranulation component is
then assumed to follow the same scalings: \citet{Corsaro2017} state that it scales with surface gravity and temperature like granulation, because
it represents a reorganization of granulation at larger scales and so originates from the same envelope convective motions. This implicitly assumes the
two components represent the same convective process at different scales, responding identically to the parameters that govern convection.

Red giants are a favorable population for testing this assumption. They span roughly two orders of magnitude in surface gravity, shifting granulation and mesogranulation to well-separated frequencies below $\nu_{\rm max}$, and they cover a range of $T_{\rm eff}$ and [Fe/H] at each gravity. This allows the dependence of each component on these fundamental stellar parameters to be tested largely independently. If granulation and mesogranulation share a common physical origin, their dependence on these parameters should take the same functional form; if they arise from convective motions at different depths, governed by different physics, there is no reason for such a connection.

Effective temperature ($T_{\rm eff}$), surface gravity($\log g$), and metallicity([Fe/H]) are the natural test parameters: $\log g$ sets the pressure scale height and, through it, convective cell size and turnover timescale \citep{Kjeldsen1995}; $T_{\rm eff}$ governs the opacity-dependent efficiency of convective energy transport \citep{Tremblay2013,Sreenivas2025}; and [Fe/H] modulates opacity through line blanketing. These same parameters, combined with $\nu_{\rm max}$, underpin asteroseismic mass, radius, and age determinations achieving relative uncertainties of order a few percent for large red giant ensembles \citep{Anders2017}; any unmodeled dependence of the background on these parameters risks biasing these quantities.

\section{Sample and Data} \label{sec: sample and data}

Our sample combines existing catalogs of red giants with measured $\nu_{\rm max}$ and evolutionary-state classifications \citep{Yu2018,Vrard2025}, yielding $\sim$ 19{,}000 stars spanning the red giant branch (RGB) and red clump (RC). RC stars comprise two sub-populations: primary red clump (pRC) stars, descendants of low-mass RGB stars ($\lesssim$ 2M$_{\odot}$) igniting helium via the degenerate core helium flash, and secondary red clump (sRC) stars ($>$ 2M$_{\odot}$), descendants of more massive stars igniting helium non-degenerately \citep{Bedding2011,Stello2013}. Because RGB, pRC, and sRC stars of similar $T_{\rm eff}$, $\log g$, and [Fe/H] ranges can differ in mass and hence in radius, we do not fit them separately. Instead we pool them and use the evolutionary phase as a diagnostic: if $T_{\rm eff}$, $\log g$, and [Fe/H], together with the radius correction of Equation~\ref{eq:amplitude_int}, fully describe the convective background, the residuals should show no offsets between phases, whereas an offset would point to a dependence on mass or evolutionary state not captured by the model. We examine the phase-resolved residuals in Sections~\ref{sec:total sample} and~\ref{sec:gp_results}.

We cross-match this sample with APOGEE \citep{Apogee} to obtain spectroscopic $T_{\rm eff}$, $\log g$, and [Fe/H], excluding stars flagged \texttt{STAR\_BAD} or \texttt{STAR\_WARN}. After quality cuts, our final sample comprises 9670 red giants (4560 RGB, 4325 pRC, 785 sRC), spanning a wide range in $T_{\rm eff}$ and $\log g$ (Figure~\ref{fig:1}, Appendix~\ref{app:sample}) and stellar mass (Figure~\ref{fig:mass distribution}, Appendix~\ref{app:sample}). RGB and pRC stars are concentrated near $\sim$1.2--1.4~M$_\odot$, while sRC stars show a broader distribution with a high-mass tail extending beyond 3~M$_\odot$. Our sample yields 84.6\% pRC and 15.4\% sRC among clump stars, closely matching the LAMOST DR7 baseline of 89.1\%/10.9\% \citep{Zhou2022}.

\section{Methods} \label{sec:model PSD}

\subsection{Background Model and Fitting}

We model the red giant power spectral density as the sum of an activity term, two super-Lorentzian convective components (granulation and mesogranulation), a Gaussian oscillation envelope, and white noise, following the background model described in \citet{Kallinger2014}:
\begin{equation}
P_{\rm bkg}(\nu) = \frac{2\pi a_1^2/b_1}{1+(\nu/b_1)^2} + \eta^2(\nu)\sum_{i=2}^{3}\frac{\zeta a_i^2/b_i}{1+(\nu/b_i)^4} + W
\label{eq:background}
\end{equation}
where $a_i$, $b_i$ are the characteristic amplitude and frequency of component $i$ ($i=2$: mesogranulation, $i=3$: granulation), $\eta(\nu)$ is the Kepler apodization function, and $W$ is the white-noise floor.

\begin{figure*}[!ht]
\centering
\gridline{
    \fig{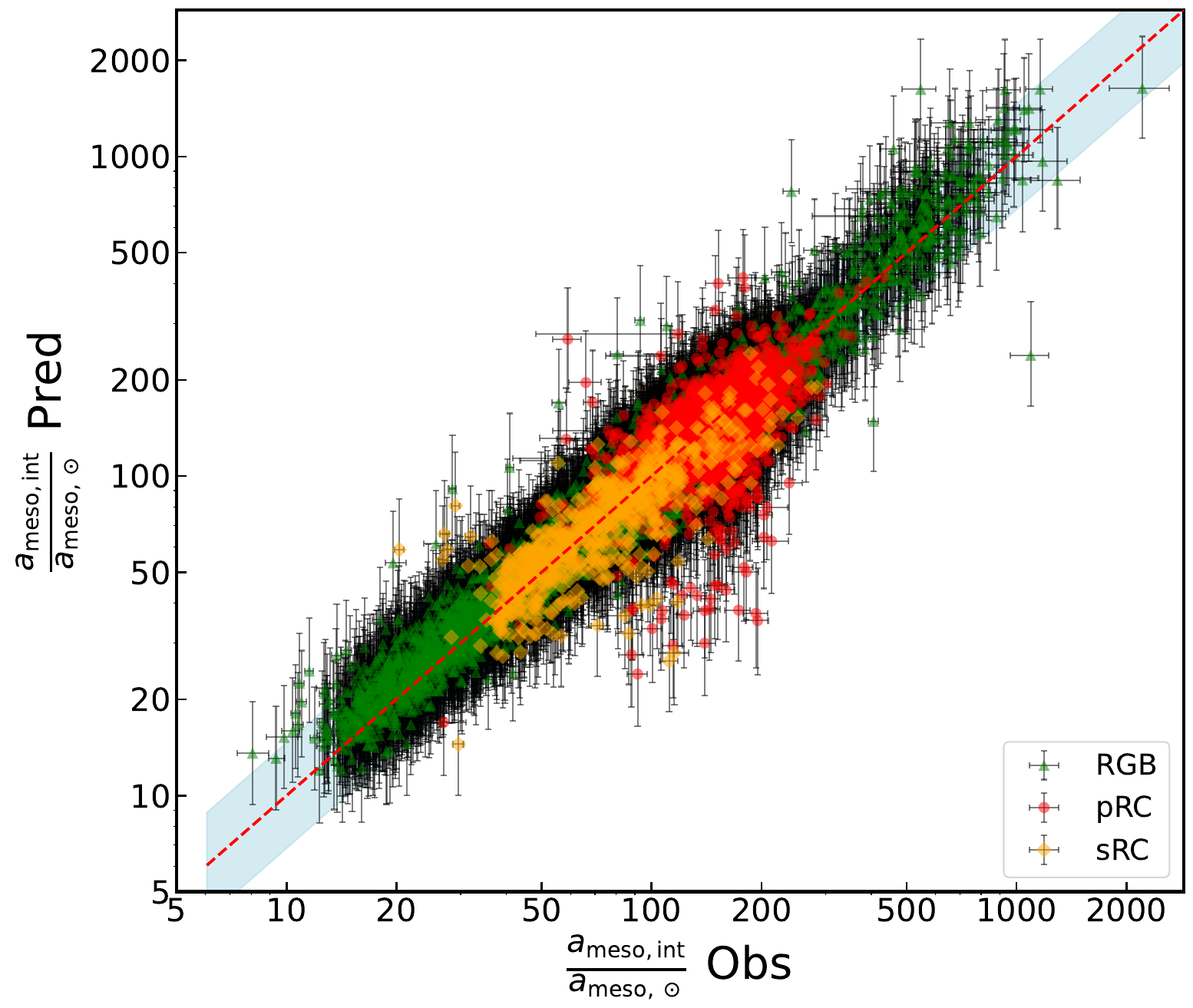}{0.49\textwidth}{(a)}
    \fig{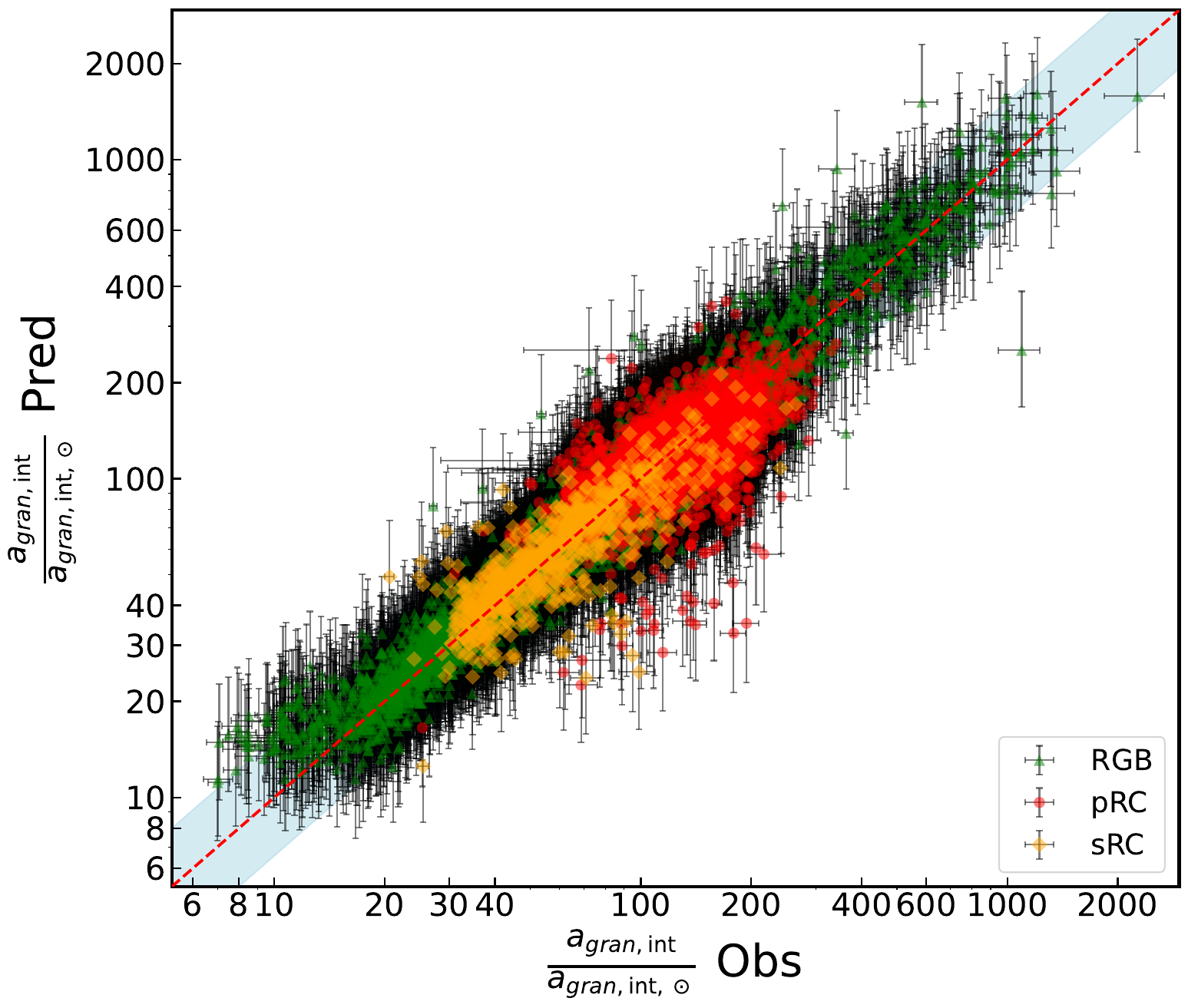}{0.49\textwidth}{(b)}
}
\gridline{
    \fig{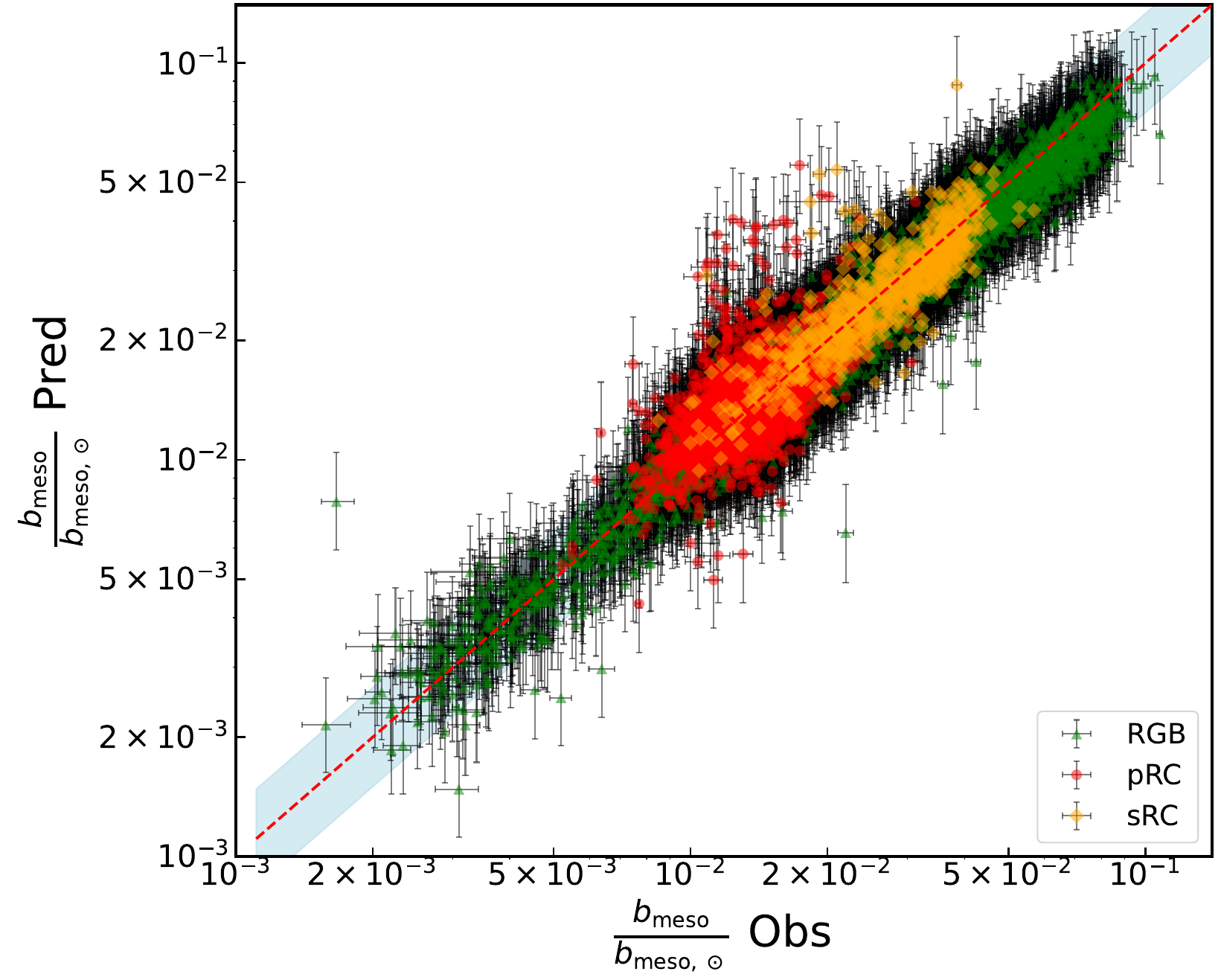}{0.49\textwidth}{(c)}
    \fig{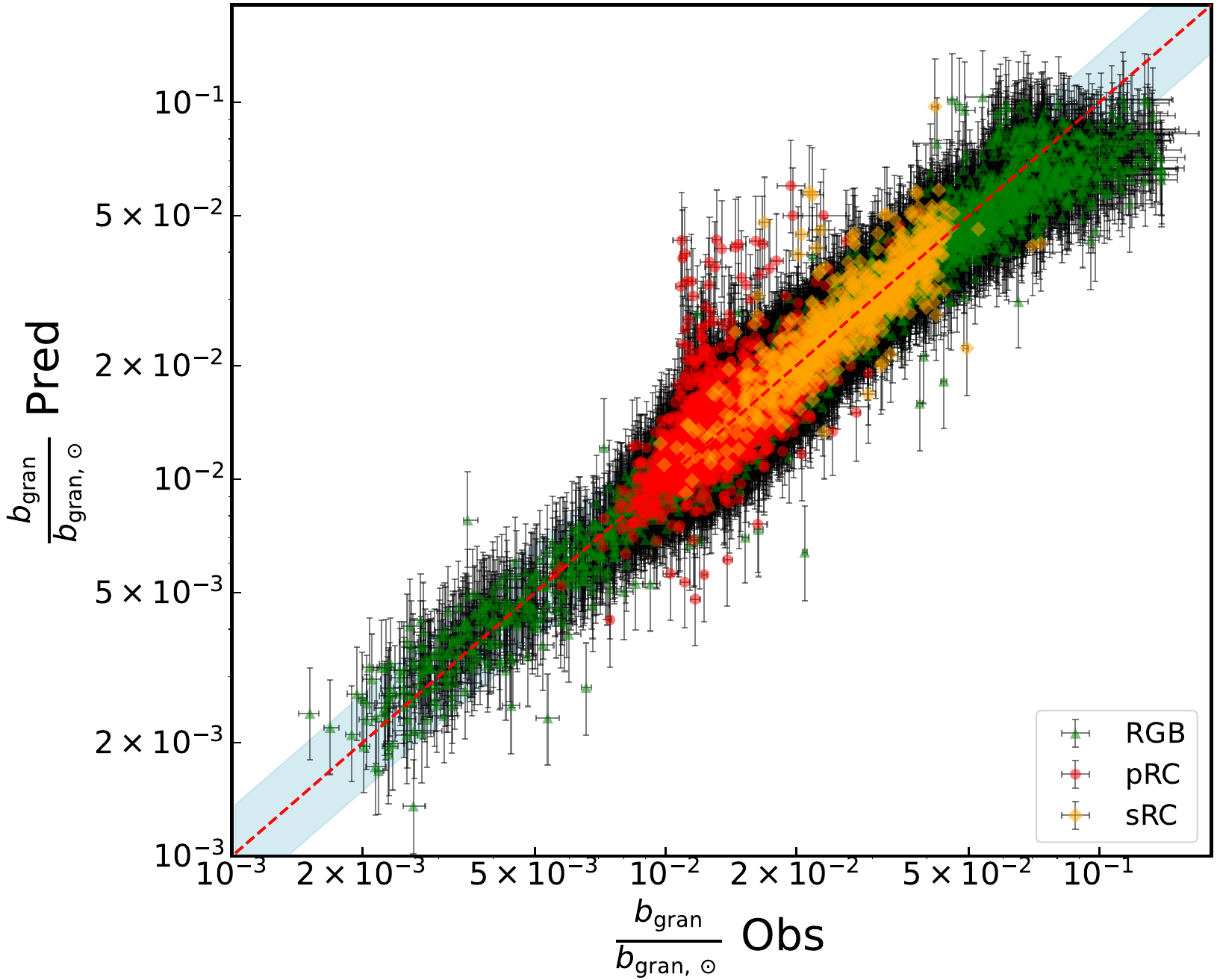}{0.49\textwidth}{(d)}
}
\caption{Predicted versus observed values of [$a_{meso}/a_{meso, \odot}, a_{gran}/a_{gran,\odot}, b_{meso}/b_{meso, \odot} / b_{gran}/b_{gran,\odot}$], using the best-fit scaling relation from Table~\ref{tab:exponents NSMC}. Points are color-coded by evolutionary phase (RGB: green triangles; pRC: red circles; sRC: orange diamonds); error bars reflect measurement uncertainties. The red dashed line marks the one-to-one relation; the shaded blue band shows the $\pm2\sigma$ scatter about this relation.}
\label{fig:obs vs predicted}
\end{figure*}

We fit this model to each star's power spectrum jointly with the oscillation envelope, using a background-fitting module of the asteroseismic peakbagging pipeline (\texttt{SPANDA} code) described in Bhattacharya et al. (in preparation), based on differential evolution initialization \citep{Storn1997} followed by Hamiltonian Monte Carlo (NUTS; \citealt{Hoffman2011}) sampling implemented in NumPyro \citep{Bingham2018,Phan2019}. We validate this pipeline against KIC~7949599, previously fit by \citet{Kallinger2014}, finding agreement within $1.4\sigma$ for all four background parameters (Appendix~\ref{app:validation}). This comparison uses the observed amplitudes, as reported by \citet{Kallinger2014}. We attribute the small remaining differences between the two sets of fitted parameters primarily to differences in how the power spectra were constructed (we use \citet[KEPSEISMIC PSDs][]{Garcia2011}) rather than to the fitting procedure itself. As a quality-control step, we exclude stars for which the mean or standard deviation of the ratio of the observed to best-fit model PSD (expected to be close to unity for a well-behaved fit) falls outside 0.98--1.02, yielding a final working sample of 9328 stars used to calibrate the scaling relations.

\begin{deluxetable}{lcccccc}
\tablecaption{Best-fit scaling relation parameters for granulation and mesogranulation \label{tab:scaling_relations}}
\tablewidth{0pt}
\tablehead{
\colhead{Parameter} & \colhead{Best Model} & \colhead{$C$} & \colhead{$\alpha$ (Teff)} & \colhead{$\beta$ (log g)} & \colhead{$\gamma$ ([Fe/H])} & \colhead{$\sigma_{\rm int}$}
}
\startdata
$a_{2,\rm int}$ (meso) & $T_{\rm eff}$, $g$, [Fe/H] & $0.911^{+0.013}_{-0.014}$ & $1.295^{+0.091}_{-0.093}$  & $-1.169^{+0.005}_{-0.005}$ & $0.350^{+0.011}_{-0.011}$& $0.190^{+0.002}_{-0.002}$ \\
$a_{3, \rm int}$ (gran) & $g$, [Fe/H]  & $0.766^{+0.012}_{-0.011}$ & \nodata  & $-1.122^{+0.004}_{-0.004}$ & $0.261^{+0.010}_{-0.010}$ & $0.211^{+0.002}_{-0.002}$ \\
$b_2$ (meso) & Teff, $g$, [Fe/H]  & $0.829^{+0.010}_{-0.010}$ & $-0.233^{+0.074}_{-0.071}$ & $0.905^{+0.004}_{-0.003}$  & $0.034^{+0.008}_{-0.009}$  & $0.144^{+0.001}_{-0.001}$ \\
$b_3$ (gran) & Teff, $g$          & $1.033^{+0.012}_{-0.013}$ & $-0.345^{+0.058}_{-0.061}$ & $0.958^{+0.003}_{-0.003}$  & \nodata& $0.148^{+0.001}_{-0.001}$ \\
\enddata
\tablecomments{Parameters correspond to Equation~\ref{eq:form}, fit using the best-supported model identified via Bayesian model comparison (Section~\ref{sec:results_scaling}). Uncertainties are the 16th--84th percentile credible interval; $\sigma_{\rm int}$ ranges 0.14--0.21 \nodata\ indicates the parameter was not included in the best-supported model.}
\label{tab:exponents NSMC}
\end{deluxetable}

\subsection{Scaling Relations and Model Selection} \label{sec:results_scaling}

We assume each of $a_2, a_3, b_2, b_3$ scales with $T_{\rm eff}$ and $\log g$ through a power law, and [Fe/H] through an exponential term:
\begin{equation}
\frac{X}{X_\odot} = C\left(\frac{T_{\rm eff}}{T_{\rm eff,\odot}}\right)^{\alpha}\left(\frac{g}{g_\odot}\right)^{\beta} e^{\gamma\,[\rm Fe/H]},
\label{eq:form}
\end{equation}
motivated by standard mixing-length expectations ($H_p \propto T_{\rm eff}/g$) and consistent with previous granulation scaling relations \citep{Mathur2011,Kallinger2014,Corsaro2017}. This form captures the intrinsic atmospheric response of each convective component, set by only local photospheric conditions ($T_{\rm eff}$, $\log g$, [Fe/H]) constructed independent of any global stellar property (mass, radius, luminosity) by design. The amplitude $a_i$ measured from the power spectra is not the intrinsic brightness contrast of a single convective cell, but the disk-integrated residual of that contrast after averaging over every independent convective cells simultaneously visible on the stellar disk. $a_i$ is thus diluted by the number of convective cells visible on the stellar disk, $N_i \sim (R/\Lambda_i)^2$, so that $a_{i,\rm obs} \sim \Theta_i\,\Lambda_i/R$ depends explicitly on the stellar radius, $R$ (see Section \ref{sec:theory} for a full discussion) Because our sample spans evolutionary sub-populations of substantially different characteristic mass at comparable surface gravity (Section \ref{sec: sample and data}), this dilution is not approximately constant across our sample. We therefore remove it explicitly using the asteroseismic radius of \citet{Yu2018},
\begin{equation}
a_{i,\rm int} = a_{i,\rm obs} \times \frac{R}{R_\odot},
\label{eq:amplitude_int}
\end{equation}
and fit $a_{i,\rm int}$, rather than $a_{i,\rm obs}$, to Equation \ref{eq:form}. These radii follow from asteroseismic scaling relations ($R \propto \nu_{\max}\,
\Delta\nu^{-2}\, T_{\rm eff}^{1/2}$) and carry uncertainties of a few per cent, which we do not propagate into $a_{i,\rm int}$. The characteristic frequency, $b_i$, depends only on local convective properties and contains no disk-integration term (Section \ref{sec:theory}); it requires no such correction and is fit to its observed value directly.
All quantities are normalized to solar reference values: $T_{\rm eff,\odot} = 5777$~K, $\log g_\odot = 4.438$ \citep{Asplund2009}, $[\mathrm{Fe/H}]_\odot = 0$, and the solar $a_{\rm i,\odot}$ and $b_{\rm i,\odot}$ from the relations of \citet{Kallinger2014} at the solar $\nu_{\max}$. These are reference values only: a different choice rescales $C$ but leaves the exponents and the model comparison unchanged \citep{Corsaro2017}. We therefore fit $C$ rather than fixing it to unity. Our sample ($\log g \lesssim 3.4$) lies $\sim$ 1dex from the Sun, so $C \neq 1$ (0.77--1.03; Table~\ref{tab:exponents NSMC}) measures how far the power
law, extrapolated to solar parameters, departs from the reference values. This may reflect systematic differences between our background fits and those used
to define the reference values, such as the construction of the power spectra (Appendix~\ref{app:validation}). The relations are not intended to reproduce solar values.
We linearize this relation and model an intrinsic scatter term, $\sigma_{\rm int}$, accounting for other astrophysical variations (e.g., evolutionary state, activity) not captured by $T_{\rm eff}$, $g$, and [Fe/H] alone; this term prevents the fit from attributing all observed dispersion to measurement uncertainty.

Rather than assuming a priori that all three parameters contribute, we treat their inclusion as a model-selection problem, fitting all seven combinations of $\{T_{\rm eff}, \log g, {\rm [Fe/H]}\}$ independently for each quantity using nested sampling \citep[UltraNest (UN);][]{Buchner2021}, and selecting the best-supported model via posterior odds ratios using the Bayes factor, with a combinatorial model prior disfavoring unnecessary complexity. Bayesian evidences for all seven candidate models are given in Appendix~\ref{app:model_comparison} (Table~\ref{tab:bayesian_model_comparison}.

\subsection{Independent Validation} \label{sec:gp_validation}

We validate the adopted functional form using two independent, model-agnostic approaches. First, we compare the predictive accuracy of each best-supported power-law model against a fully flexible Gaussian Process (GP) regressor \citep{Rasmussen2006} with an anisotropic radial basis function (RBF; squared-exponential) kernel, which assigns a separate correlation length scale to each predictor, trained on the same predictors and evaluated via root-mean-square error (RMSE) on a held-out test set (full kernel and RMSE comparison in Appendix~\ref{app:gp}, Table~\ref{tab:gp_summary}). Second, we apply symbolic regression \citep[PySR;][]{Cranmer2023} directly to each quantity, which searches the space of analytic expressions, using an algorithm that repeatedly mutates and recombines a population of candidate expressions to search for the simplest analytic expression that fits each quantity, with no functional form assumed in advance. It penalizes complexity, so that the result is the simplest equation that fits well rather than a fixed form with fitted coefficients. The regression operator can set limits on which type of functional forms can be found. Because the search is stochastic, we repeat it with five different random seeds, i.e.\ different
initializations of a random number generator, which lead to different search trajectories, and select the run with the smallest RMSE of the log-space residuals, weighted by the inverse variance of the measurement uncertainties (Appendix~\ref{app:sr}, Table~\ref{tab:sr_summary}).

\section{Results}

\subsection{Scaling Relations} \label{sec:derived scaling relation}

Table~\ref{tab:exponents NSMC} summarizes the best-fit parameters for each quantity, together with the model selected via Bayesian model comparison. For the amplitudes, the full three-parameter model is preferred for mesogranulation ($a_{2,\rm int}$), with $\alpha = 1.295$, $\beta = -1.169$, $\gamma = 0.350$; for granulation ($a_{3,\rm int}$), the best-supported model excludes $T_{\rm eff}$ entirely, retaining only $\log g$ and [Fe/H] ($\beta = -1.122$, $\gamma = 0.261$). Both amplitudes show a steep, comparable negative $\log g$ dependence ($\beta \approx -1.1$). Metallicity shows a substantial, positive exponent for both components. For the frequencies, the full model is preferred for $b_2$, and $T_{\rm eff}+\log g$ for $b_3$; both show weak, negative $T_{\rm eff}$ dependence and strong, closely matched, positive $\log g$ dependence ($\beta \approx 0.9$--$1.0$). The intrinsic scatter, $\sigma_{\rm int}$, is small throughout (0.14--0.21).

\begin{figure*}[!htbp]
\centering
\gridline{
    \fig{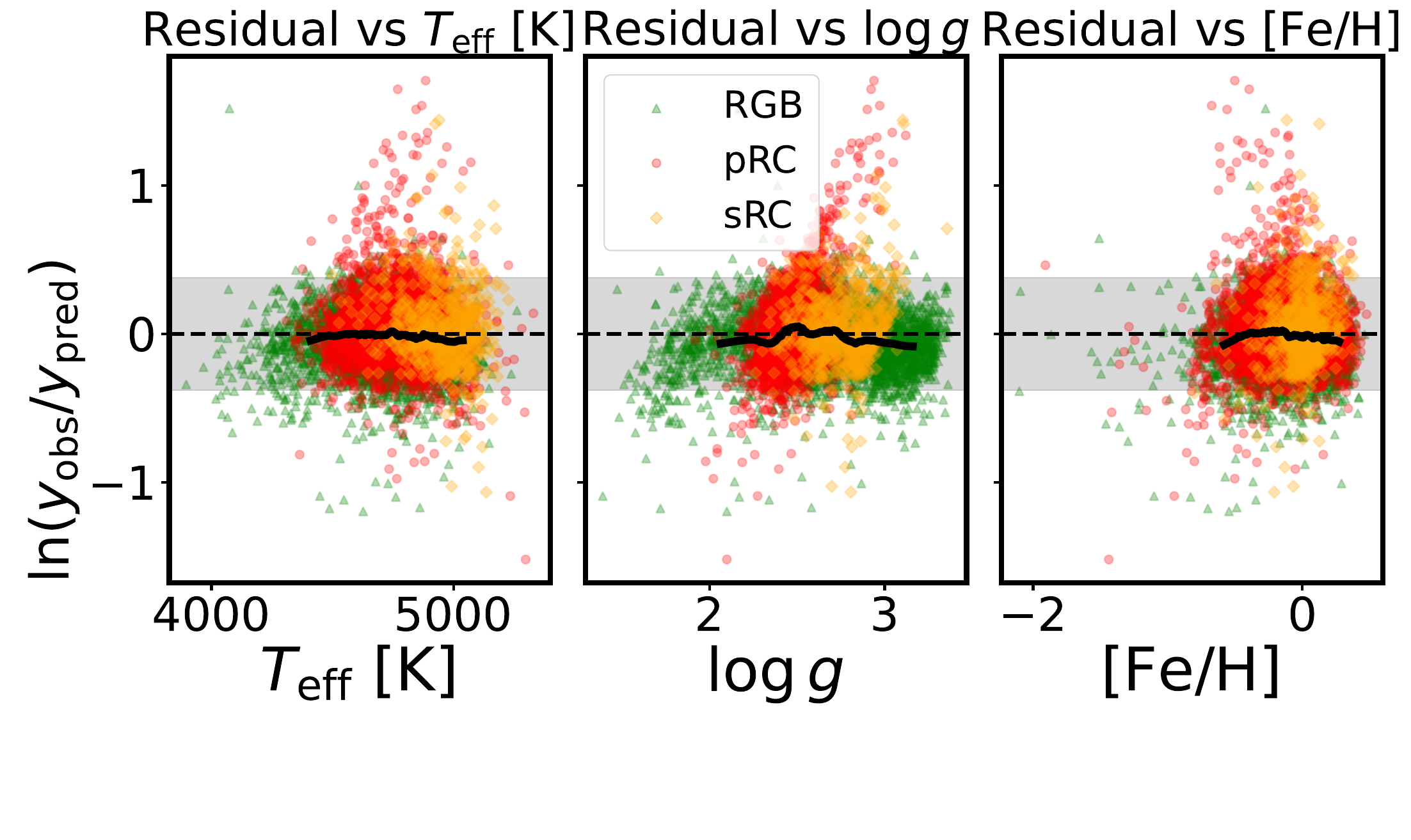}{0.46\textwidth}{(a) $a_{2, \rm int}$}
    \fig{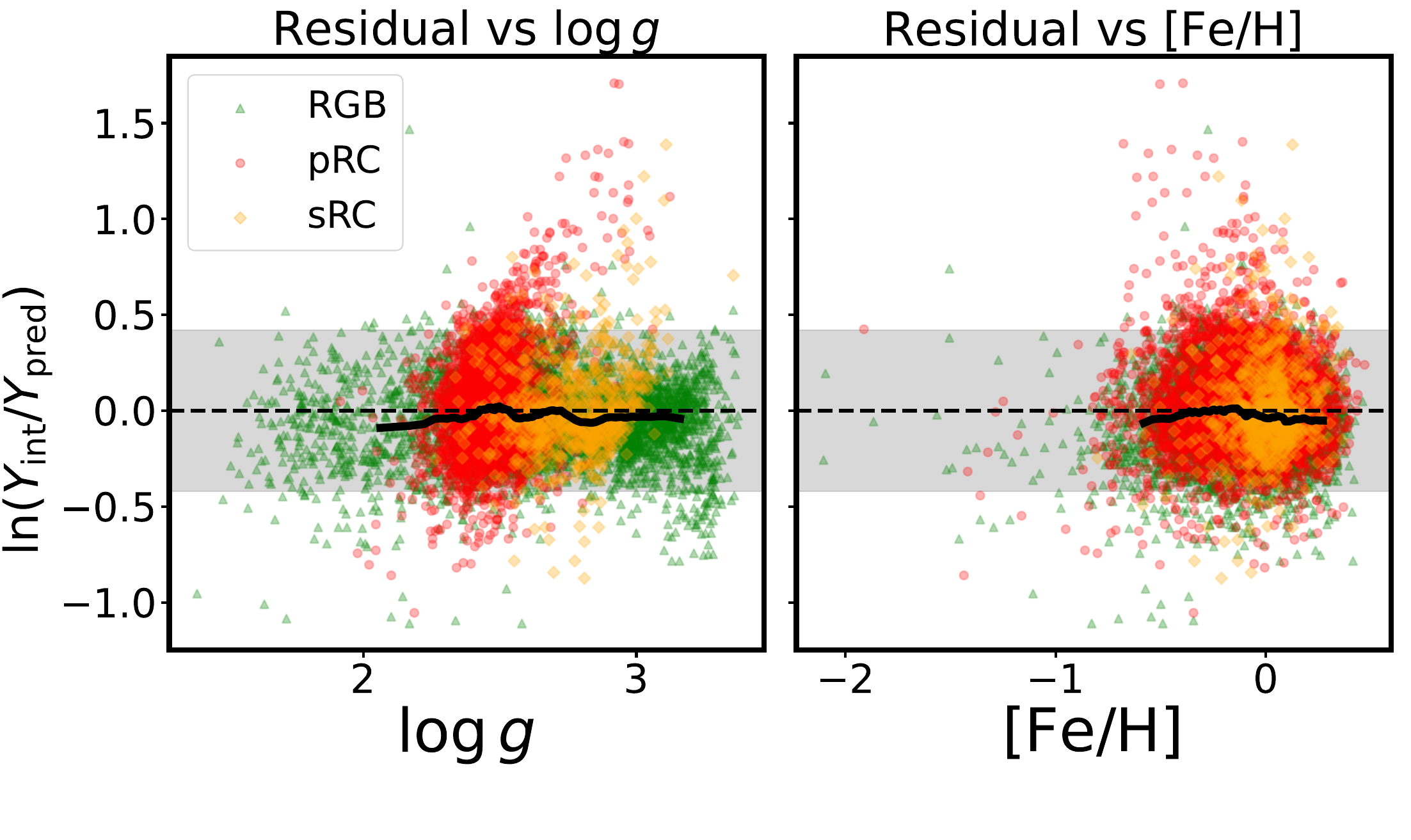}{0.46\textwidth}{(b) $a_{3, \rm int}$}}
\gridline{
    \fig{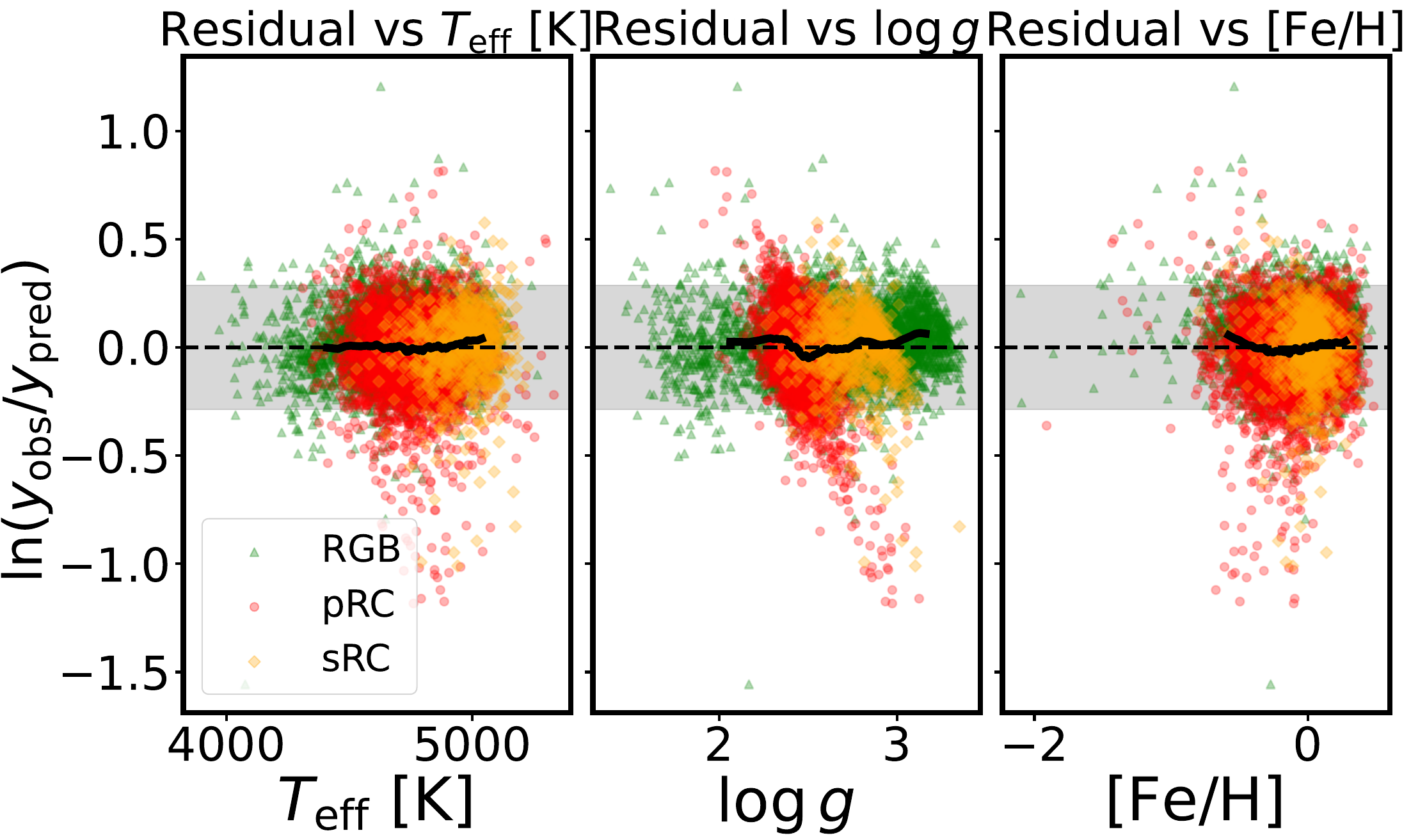}{0.45\textwidth}{(c) $b_2$}
    \fig{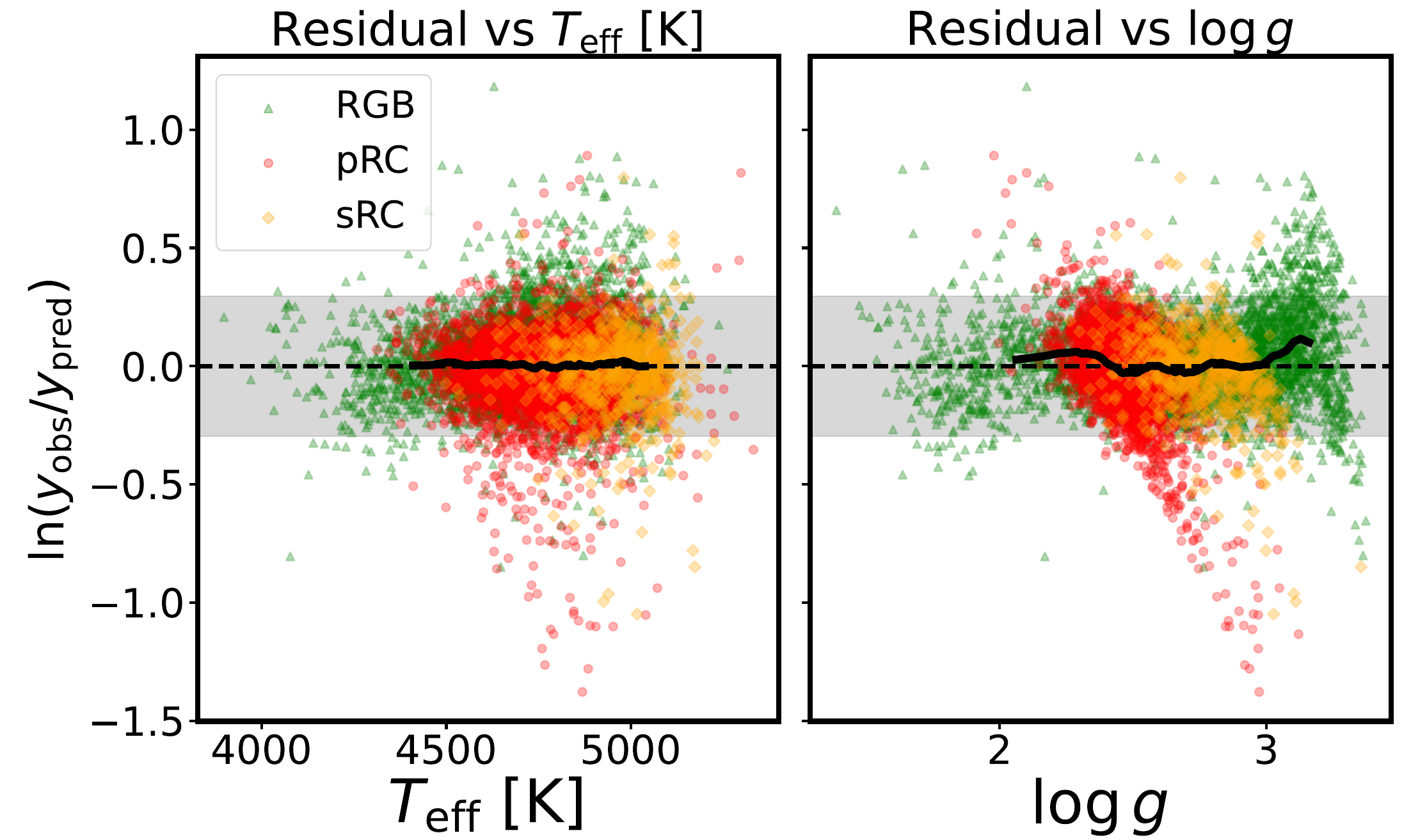}{0.45\textwidth}{(d) $b_3$}}
\gridline{
    \fig{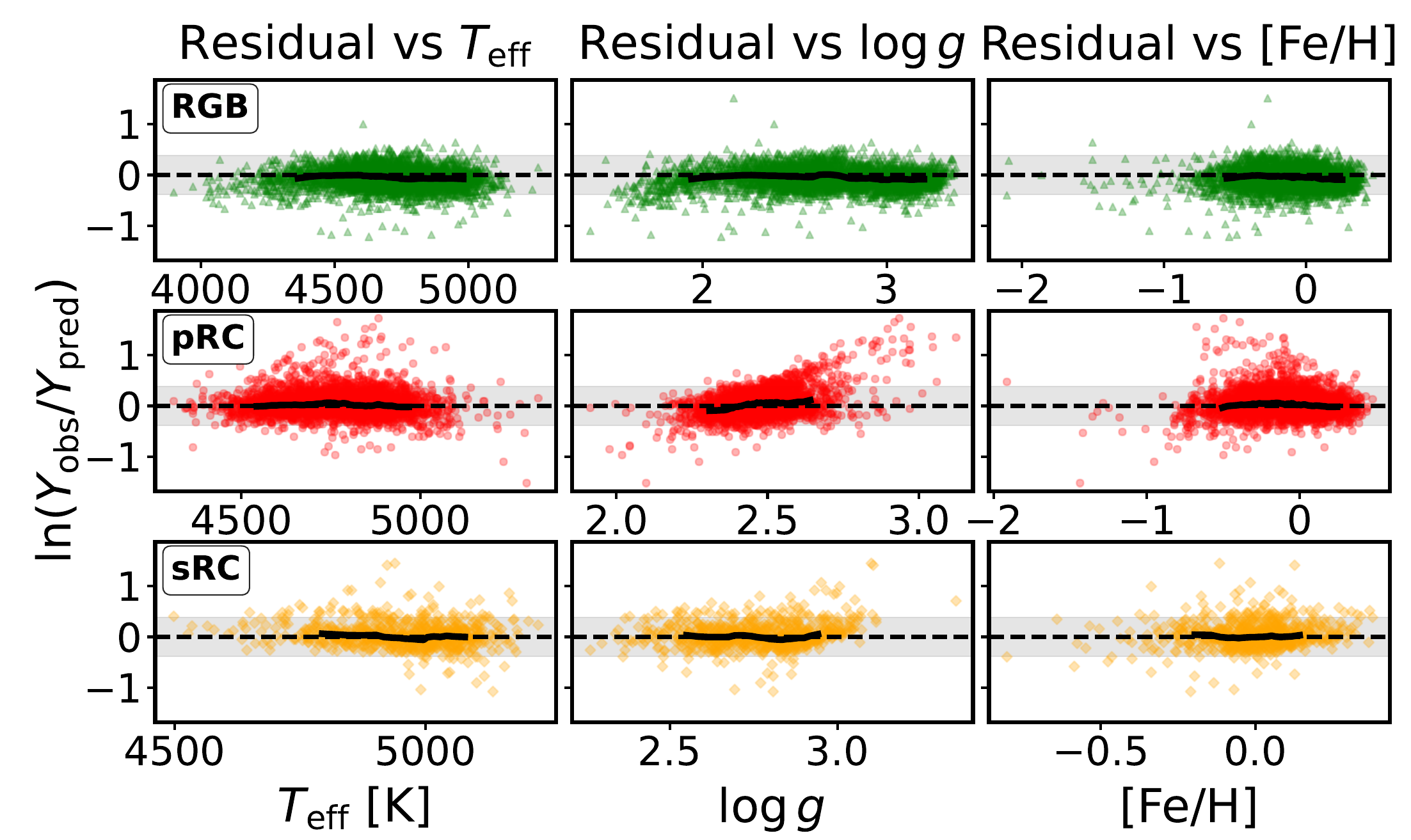}{0.45\textwidth}{(e) $a_{2, \rm int}$}
    \fig{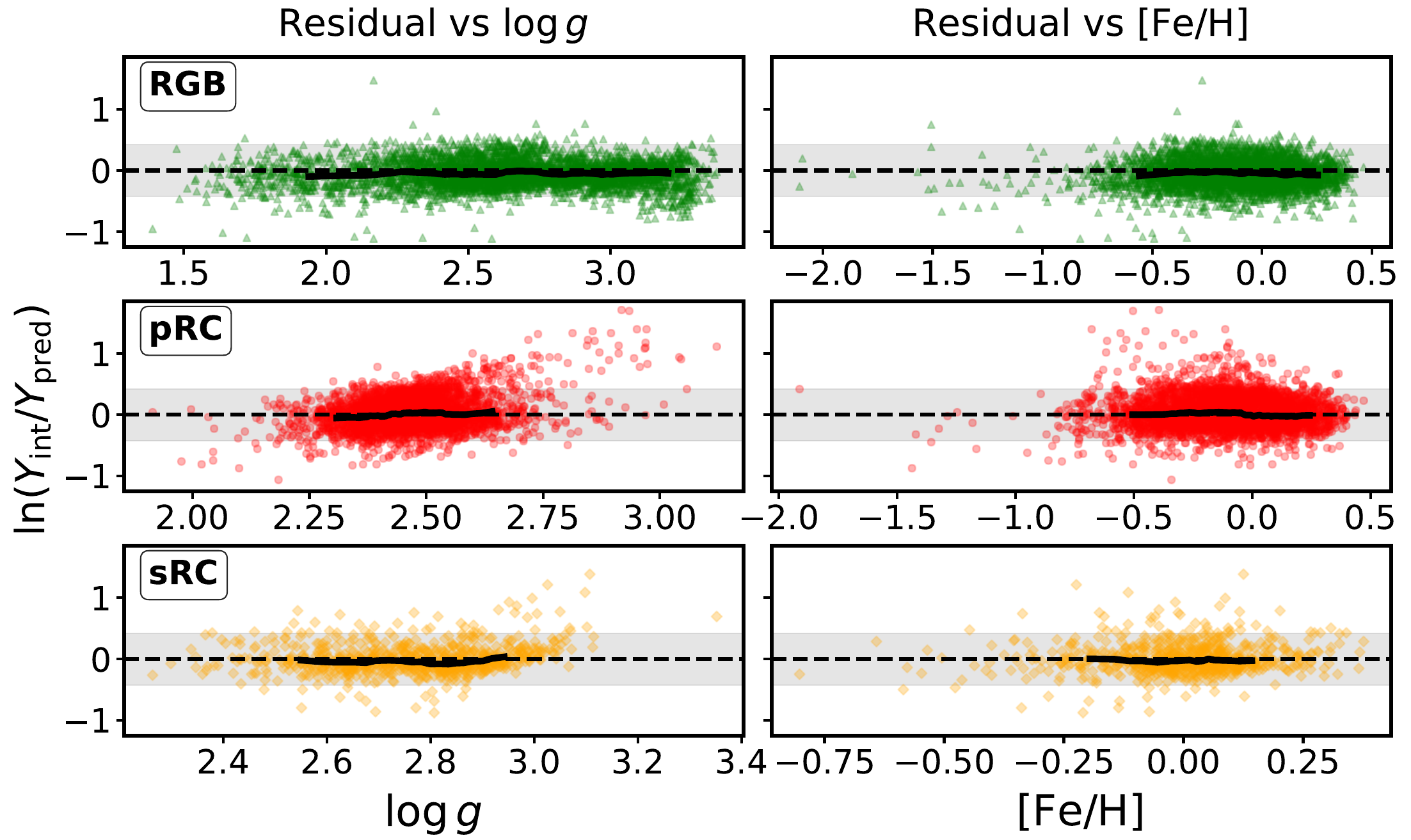}{0.45\textwidth}{(f) $a_{3, \rm int}$}}
\gridline{
    \fig{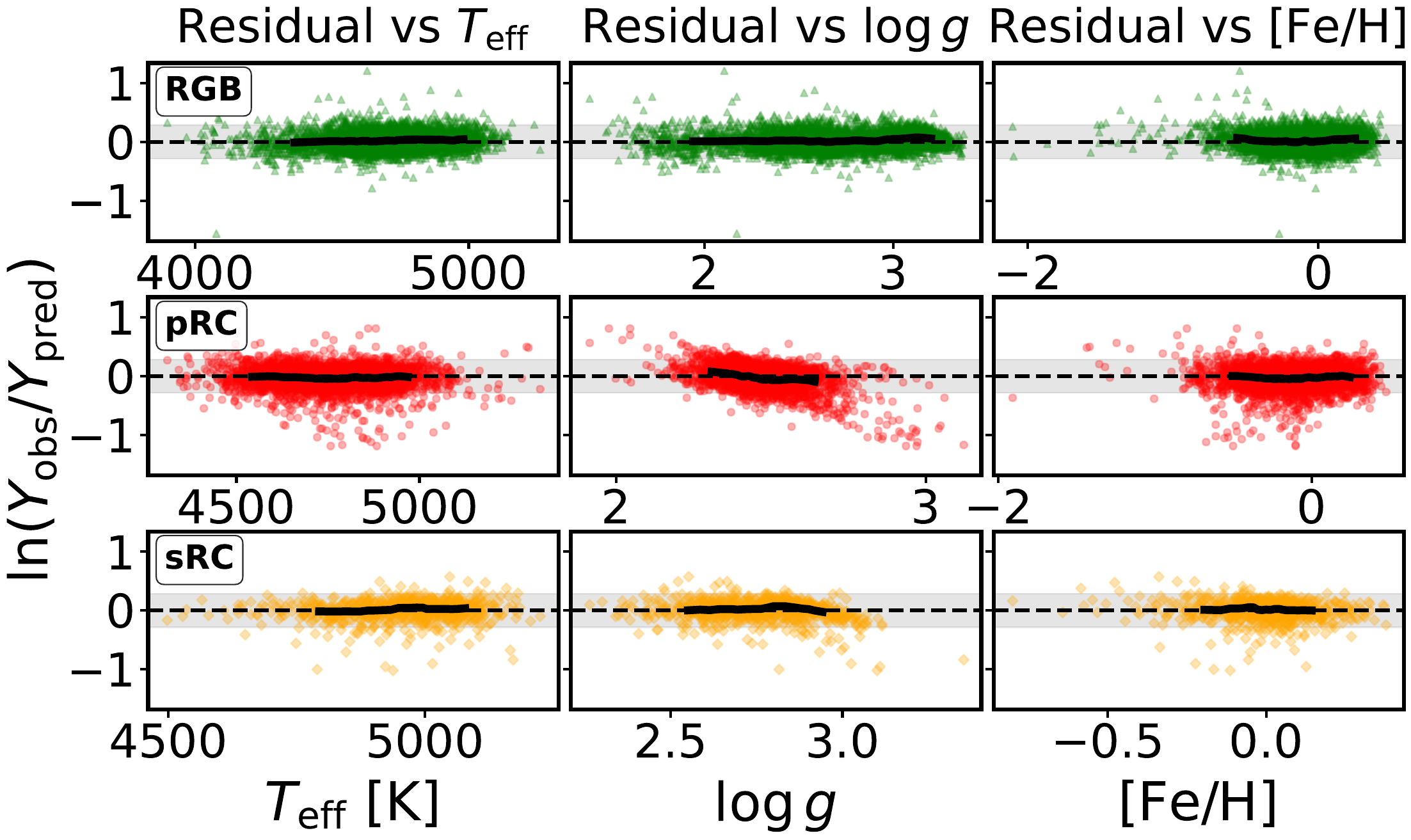}{0.45\textwidth}{(g) $b_2$}
    \fig{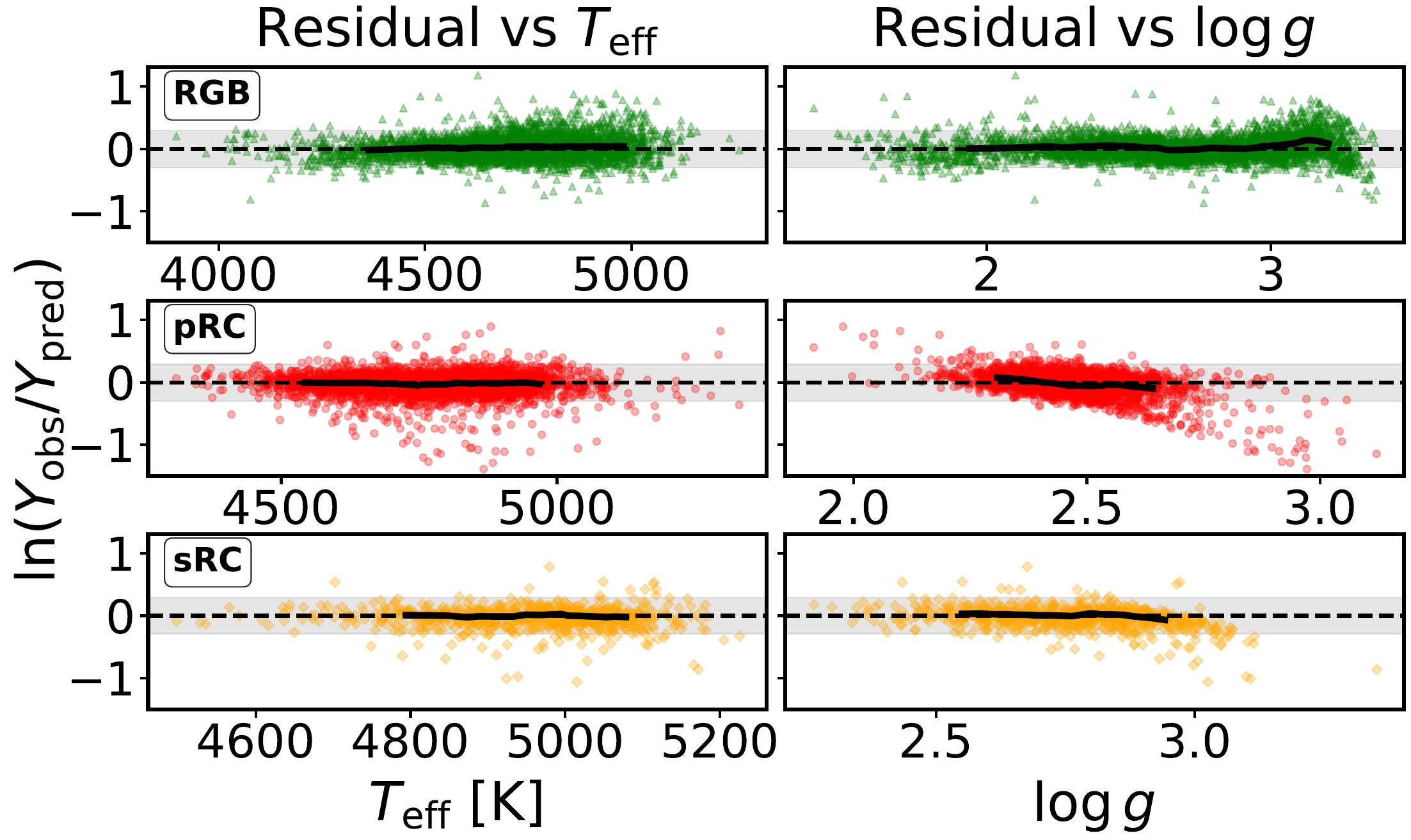}{0.45\textwidth}{(h) $b_3$}
}
\caption{Residuals of the predicted-versus-observed comparison, $\ln(Y_{\rm obs}/Y_{\rm pred})$, for $a_{2, \rm int}, a_{3, \rm int}, b_2, b_3$, as a function of $T_{\rm eff}$, $\log g$, and [Fe/H] where included in the adopted model, for the full sample. The black line shows a running median; the shaded band indicates the $\pm2\sigma$ region. Top row (a--d): full sample, color-coded by evolutionary phase. Bottom row (e--h): the same residuals separated into individual RGB (top), pRC (middle), and sRC (bottom) sub-panels, with the running median computed independently within each evolutionary phase}
\label{fig:residuals merged}
\end{figure*}

Across all four quantities, the $\log g$ dependence is closely comparable between granulation and mesogranulation, consistent with both signals being governed by a shared intrinsic atmospheric response to surface gravity, once the radius-dependent geometric dilution has been removed; by contrast, the $T_{\rm eff}$ dependence diverges substantially: mesogranulation's intrinsic amplitude retains a substantial, positive $T_{\rm eff}$ dependence, while granulation's intrinsic amplitude shows no statistically justified $T_{\rm eff}$ dependence at all; both frequencies retain weak $T_{\rm eff}$ sensitivity, comparable to one another. We discuss the physical implications of this pattern, including the role of metallicity, in Section \ref{sec:discussion}.

\subsection{Predicted versus Observed Comparison} \label{sec:total sample}

Figure~\ref{fig:obs vs predicted} compares, for each quantity, the predicted value from Table~\ref{tab:exponents NSMC} against the observed value for every star, color-coded by evolutionary phase. For the amplitudes, the quantities shown are the intrinsic values $a_{i,\rm int}$; for the frequencies, they are the observed values $b_i$. The great majority of stars fall within the $\pm2\sigma$ band; only 5--9\% fall outside, consistent with expected statistical scatter.

We examine the residuals, $\ln(Y_{\rm int}/Y_{\rm pred})$ for the amplitudes and $\ln(Y_{\rm obs}/Y_{\rm pred})$ for the frequencies, as a function of the stellar parameters (Figure~\ref{fig:residuals merged}). For both amplitudes $a_{i,\rm int}$, and frequencies $b_i$, the running median (computed in overlapping windows of 600 stars, offset by 100 stars) stays within $\approx \pm$0.1 of zero across the full range of the corresponding best-fit model predictors. When resolved by evolutionary phase, the RGB, pRC, and sRC amplitude residuals are each flat with median offsets within $\approx 5\%$. For the frequencies, the pRC median declines mildly with $\log g$, from $\approx$ +0.05 at $\log g \approx$ 2.3 to $\approx$ -0.1 at $\log g \approx$ 2.65, with a similar but weaker decline for sRC above $\log g \approx$ 2.8. The RGB frequency residuals remain flat, apart from a slight rise to $\approx$ +0.1 at $\log g \gtrsim$ 3.1. Because RGB, pRC, and sRC occupy overlapping but distinct regions of $T_{\rm eff}-\log g$ space, this population mixing also explains the small wave-like structure ($\approx$ 0.05) seen when all stars are combined.

The dominant feature in the residuals is a small number of pRC stars (and a few sRC stars) at $\log g \gtrsim$ 2.7. These stars have large positive amplitude residuals($\sim$+1.7), and negative frequency residuals( $\sim$ -1.4). These stars lie well above the red-clump locus ($\log g \approx$ 2.3–2.7), RGB stars at the same log g show no comparable tail, and the opposite signs of the amplitude and frequency residuals are those expected from an overestimated surface gravity, given the signs of $\beta$ in Table \ref{tab:exponents NSMC}. For the 108 pRC stars with spectroscopic $\log g \gtrsim$ 2.7, the seismic $\log g$ derived from $\nu_{max}$ and $T_{\rm eff}$ is lower than the APOGEE value by a median of 0.18 dex (up to 0.65 dex), enough to account for the residuals. We therefore attribute these outliers to overestimated spectroscopic surface gravities, rather than to the atmospheric response of the convective components. These stars are too few to shift the running medians.

\begin{deluxetable*}{llrrccccc}[!htb]
\tabletypesize{\scriptsize}
\tablewidth{0pt}
\tablecaption{Phase-resolved comparison of the GP and parametric predictions on the withheld test sample. \label{tab:gp_phase}}
\tablehead{
\colhead{Quantity} & \colhead{Phase} & \colhead{$N_{\rm test}$} & \colhead{$\widetilde{M}/M_\odot$} & \colhead{Offset$_{\rm par}$} & \colhead{Offset$_{\rm GP}$} & \colhead{RMSE$_{\rm par}$} & \colhead{RMSE$_{\rm GP}$} & \colhead{Improvement}
}
\startdata
$a_{2,\rm int}$ & RGB & 918 & 1.30 & -4.22\% & -3.27\% & 0.195 & 0.191 &  2.32\% \\
          & pRC & 807 & 1.25 & +1.55\% & +1.52\% & 0.242 & 0.225 &  7.09\% \\
          & sRC & 141 & 2.28 & -1.59\% & -4.71\% & 0.284 & 0.294 & -3.59\% \\
\hline
$a_{3,\rm int}$ & RGB & 918 & 1.30 & -4.16\% & -3.87\% & 0.219 & 0.218 &  0.57\% \\
          & pRC & 807 & 1.25 & -0.33\% & -0.99\% & 0.268 & 0.263 &  1.94\% \\
          & sRC & 141 & 2.28 & -4.74\% & -5.76\% & 0.279 & 0.289 & -3.51\% \\
\hline
$b_2$ & RGB & 918 & 1.30 & $+3.14\%$ & $+3.07\%$ & 0.154 & 0.149 &  $3.25\%$ \\
      & pRC & 807 & 1.25 & $-2.88\%$ & $-2.64\%$ & 0.188 & 0.180 &  $4.26\%$ \\
      & sRC & 141 & 2.28 & $+0.49\%$ & $+0.59\%$ & 0.219 & 0.220 & $-0.45\%$ \\
\hline
$b_3$ & RGB & 918 & 1.30 & $+2.89\%$ & $+3.92\%$ & 0.181 & 0.174 &  $3.87\%$ \\
      & pRC & 807 & 1.25 & $-1.77\%$ & $-2.06\%$ & 0.177 & 0.169 &  $4.73\%$ \\
      & sRC & 141 & 2.28 & $-0.52\%$ & $+2.37\%$ & 0.208 & 0.209 & $-0.48\%$ \\
\enddata
\tablecomments{Residual offsets in per cent, $100[\exp(\widetilde{r})-1]$, where $\widetilde{r}$ is the median of $r=\ln(X_{\rm obs}/X_\odot)-\ln(X_{\rm pred}/X_\odot)$. Positive offsets indicate the observed value exceeds the prediction. Improvement is $100(\mathrm{RMSE}_{\rm par}-\mathrm{RMSE}_{\rm GP})/\mathrm{RMSE}_{\rm par}$.}
\end{deluxetable*}

\begin{figure*}[!htb]
\centering
\includegraphics[width=0.9\textwidth]{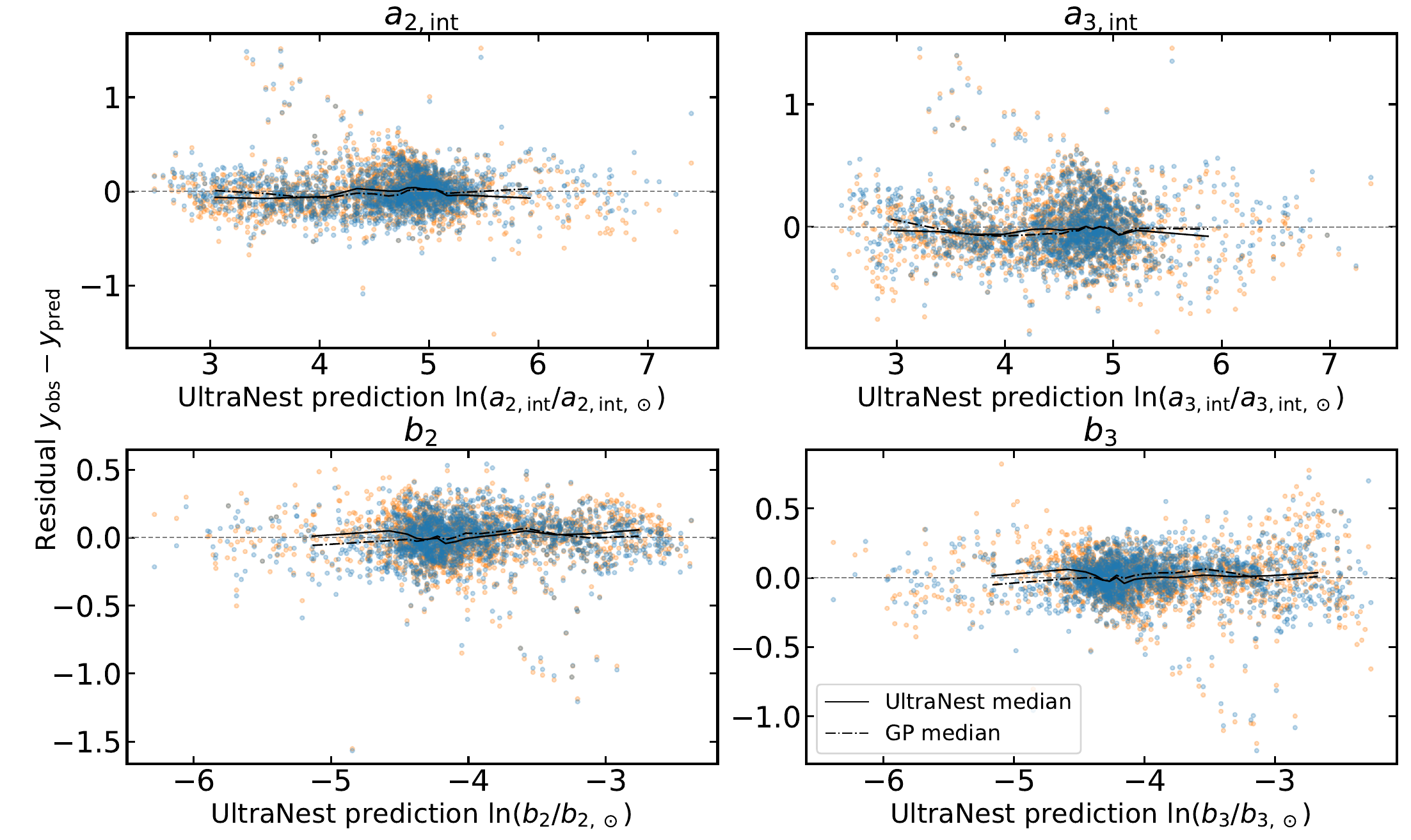}
\caption{Residuals of the parametric and GP predictions against the parametric prediction. Orange and blue points denote the parametric and GP predictions; the black solid and dash-dotted lines show the respective running medians; the horizontal dashed line marks zero residual.}
\label{fig:gp_obspred}
\end{figure*}

\subsection{Validation Results} \label{sec:gp_results}
The GP reduces the log-space RMSE of the entire sample relative to the parametric relations by 3.9\% and 0.9\% for $a_{2,\rm int}$ and $a_{3,\rm int}$, and by 3.4\% and 3.8\% for $b_2$ and $b_3$ (Appendix \ref{app:gp}, Table~\ref{tab:gp_summary}), indicating that the power-law forms capture nearly all of the predictive structure in the data. For $a_{3,\rm int}$ the correction is nearly uniform. For $a_{2,\rm int}$ it is a mild positive ($\approx +0.05$) at $\log g \approx 2.6$--$2.9$ and a negative at the lowest $\log g$. The frequency corrections are somewhat
larger and more coherent, reaching $\sim +0.15$ at $\log g \lesssim 2.3$ for $b_2$ and at $\log g \gtrsim 3$ for $b_3$, suggesting mild curvature in the
$\log g$ dependence that a single power law does not capture.

The phase-resolved comparison in Table~\ref{tab:gp_phase} shows that the GP improves the
amplitude predictions only modestly, and not uniformly across phases. For $a_{2,\rm int}$ the log-space RMSE decreases by 2.3\% for RGB and 7.1\% for pRC stars, and for $a_{3,\rm int}$ by 0.6\% and 1.9\%, whereas for sRC stars it increases by 3.6\% and 3.5\%. The median offsets of the parametric relations are -4.2\%, +1.5\%, and -1.6\% (RGB, pRC, sRC) for $a_{2,\rm int}$ and -4.2\%, -0.3\%, and -4.7\% for $a_{3,\rm int}$, so no phase departs from the relation by more than $\approx 5\%$. The GP does not consistently move these offsets closer to zero; for sRC stars it moves them away, to -4.7\% and -5.8\%.

Evolutionary phase and mass were not supplied as GP predictors. These comparisons therefore assess how well relations based on the atmospheric parameters describe each population. For $b_2$ and $b_3$, the GP reduces the RMSE by 3.3--4.7\% for RGB and pRC stars, while increasing it by less than 0.5\% for sRC stars; the median offsets, within $\pm 4\%$ for both methods, do
not consistently move closer to zero. The pooled power-law relations therefore describe RGB, pRC, and sRC stars alike to within a few per cent.

Symbolic regression, applied independently to each quantity with no assumed functional form, recovers the same structure as the Bayesian analysis. For $a_{3,int}$ it selects an equation with no $T_{\rm eff}$ dependence, depending only on $\log g$ and [Fe/H] ($\beta$ = -1.128, $\gamma$ = 0.269, versus $\beta$ = -1.122, $\gamma$ = 0.261 from the Bayesian fit), while for $a_{2,int}$ it retains a positive $T_{\rm eff}$ term ($\alpha$ = 1.0, $\beta$ = -1.198, $\gamma$ = 0.273, versus $\alpha$ = 1.295, $\beta$ = -1.169, $\gamma$ = 0.350). For the frequencies, it finds $\alpha$ = -0.202, $\beta$ = 0.924 for $b_2$ and $\alpha$ = -0.299, $\beta$ = 0.937 for $b_3$, close to the Bayesian values. Metallicity enters the selected equations for both amplitudes with positive exponents, but not for either frequency; the small Bayesian $b_2$ exponent ($\gamma$ = 0.034) is not selected. Full results are given in Appendix \ref{app:sr} (Table \ref{tab:sr_summary}), and a direct visual comparison of all three methods against the observed values is shown in Figure \ref{fig:threeway_comparison}.

\section{Discussion}\label{sec:discussion}

\subsection{A theoretical framework for the scaling relations} \label{sec:theory}

The scaling relations of Section~\ref{sec:derived scaling relation} are empirical; we assess whether they are consistent with a simple physical picture of convective granulation. Writing the characteristic length and velocity scale of component $i$ as $\Lambda_i = \eta_i H_p$ and $v_i = \mathcal{M}_i c_s$, with $H_p \propto T_{\rm eff}/g$ the pressure scale height, $\eta_i$ the cell size in units of $H_p$, $\mathcal{M}_i$ the local Mach number, and $c_s \propto T_{\rm eff}^{1/2}$, the characteristic frequency follows $b_i \sim v_i/\Lambda_i \propto (\mathcal{M}_i/\eta_i)\,g\,T_{\rm eff}^{-1/2}$ --- the same scaling underlying the granulation-timescale treatment of \citet{Samadi2013a,Samadi2013b}. If $\eta_i$ and $\mathcal{M}_i$ are approximately constant, this predicts $\alpha_b \simeq -0.5$, $\beta_b \simeq +1$.

The disk-integrated amplitude is diluted by the number of independent convective structures visible on the disk, $N_i \sim (R/\Lambda_i)^2$, giving $a_{i,\rm obs} \sim \Theta_i/\sqrt{N_i} \sim \Theta_i (\Lambda_i/R)$ where $\Theta_i$ is the intrinsic intensity contrast \citep{Kallinger2014, Corsaro2017}. Because this dilution depends on the global stellar radius, which differs between stars of the same surface gravity, we remove it explicitly through Equation \ref{eq:amplitude_int}. The intrinsic amplitude, $a_{i,\rm int} \sim \Theta_i \Lambda_i = \Theta_i \eta_i H_p \sim \Theta_i \eta_i T_{\rm eff}/g$,  contains the cell size as well as the contrast, and is then set by local atmospheric conditions alone. If $\Theta_i$ and $\eta_i$ are approximately constant, this predicts $\alpha_a \approx$ 1, $\beta_a \approx$ -1.

Our measured $\log g$ exponents agree closely with both predictions ($\beta_b$ = 0.905–0.958,  $\beta_a$ = -1.122 to -1.169) and are nearly identical between granulation and mesogranulation, as expected if both length scales are set by the same pressure scale height, independent of a component's depth or driving mechanism. The remaining deviations, $\lesssim$ 0.17, would require a weak $g$ dependence of $\mathcal{M}_i, \eta_i$ or $\Theta_i$, which we do not pursue here.

The $T_{\rm eff}$ exponents are less uniform. The mesogranulation amplitude exponent ($\alpha$ = 1.295) is somewhat above the constant-$\eta_i, \Theta_i$ expectation of +1, whereas granulation's amplitude shows no statistically justified $T_{\rm eff}$ dependence ($\alpha_a$ = 0). Both frequency exponents (-0.233 and -0.345) are less negative than the -0.5 expected for constant $\mathcal{M}_i$ and $\eta_i$. These departures require genuine $\mathcal{M}_i(T_{\rm eff})$ and $\eta_i(T_{\rm eff})$ dependencies, which we examine below.

\subsection{Temperature dependence and the amplitude--frequency asymmetry} \label{sec: Teff discussion}

The clearest difference between the two components is the response of their intrinsic amplitudes to $T_{\rm eff}$. For granulation, the model without $T_{\rm eff}$ is preferred (Bayes factor $\approx$ 20), and when $T_{\rm eff}$ is included its exponent, $\alpha$ = -0.06 $\pm$ 0.10, is consistent with zero; the Gaussian Process and symbolic regression agree (Section \ref{sec:gp_results}). The mesogranulation amplitude instead rises with $T_{\rm eff}$ ($\alpha$ = 1.295 $\pm$ 0.092), $\approx$ 10$\sigma$ above granulation, while the frequency exponents agree within $\sim$1.2$\sigma$ (-0.233 $\pm$ 0.07 and -0.345 $\pm$ 0.06).

From Section {\ref{sec:theory}}, $a_{i,\rm int} \sim \Theta_i \eta_i T_{\rm eff}/g$. $\Theta_i = (\Delta B/B)_{\rm rms}$ is the intrinsic intensity contrast of a cell, i.e. the fractional brightness difference between rising and sinking gas in the Kepler band. This contrast depends on the temperature difference of the flow, $\Theta_{\rm rms,i} = (\Delta T/T)_{\rm rms}$ \citep{Samadi2013b}, through the Planck function $B_{\lambda}(T)$. Two temperature effects enter both amplitude components ($a_{2,\rm int}, a_{3,\rm int}$).
\begin{enumerate}
\item $a_{i,\rm int}$ alone grows as $T_{\rm eff}^1$. 
\item $\Theta_i = (hc/\lambda k_BT) \Theta_{\rm rms} = (hc/\lambda k_BT) (\Delta T/T)_{\rm rms}$ in the Wien regime. Kepler observes between $\approx$ 400 and 850 nm, centred near 600 nm, where at our samples' median $T_{\rm eff} \approx$ 4750 K the Planck function is in the Wien regime ($hc/\lambda k_BT \approx$ 5 at 600 nm; the Wien approximation is accurate to $\lesssim$ 3\% even at 850 nm). The factor $hc/\lambda k_BT$ scales as $T_{\rm eff}^{-1}$ (-0.9 to -1.0 with the exact Planck function across the bandpass).
\end{enumerate}
So for a given $\Theta_{\rm rms}$ the two effects nearly cancel and leave a net dependence of $\approx T_{\rm eff}^0$. Any remaining $T_{\rm eff}$ dependence must come from the dimensionless convective properties,  $\Theta_{\rm rms,i}$ and $\eta_i$.

Following \citet{Tremblay2013}, $\Theta_{\rm rms,i} \propto \mathcal{M}_i^2$. Writing $\mathcal{M}_i \propto T_{\rm eff}^{m_i}$, $\eta_i \propto T_{\rm eff}^{\ell_i}$, the resulting exponents are $\alpha_{a,i} = 1+2m_i+\ell_i$, $\alpha_{b,i} = -\tfrac{1}{2}+m_i-\ell_i$. Solving for our measured exponents from Table \ref{tab:exponents NSMC} gives $m_2 \simeq 0.52$, $\ell_2 \simeq 0.25$ for mesogranulation, and $m_3 \simeq 0.05$, $\ell_3 \simeq -0.10$ for granulation. This decomposition is not a unique, independent measurement. It assumes
$\Theta_{{\rm rms},i}\propto\mathcal{M}_i^2$, so any further $T_{\rm eff}$ dependence of the contrast, for instance through the efficiency of radiative
cooling at the surface, is absorbed into $m_i$ and $\ell_i$. It does show that the four exponents are mutually consistent within this framework, and it
identifies whether the Mach number or the cell size drives each component's $T_{\rm eff}$ sensitivity.

Granulation's Mach number and cell size therefore vary little with $T_{\rm eff}$, whereas mesogranulation's both increase. The two are physically linked: in 3D simulations the granule size in units of $H_p$ grows with the Mach number, from $\approx$3 at $\mathcal{M \approx}$ 0.1 to $\approx$10 at $\mathcal{M \approx}$ 1, because mass conservation ties the horizontal--to--vertical velocity ratio to the cell width \citep{Tremblay2013}, which implies $\ell_i \approx$ (0.3–0.5) $m_i$; the ratio $\ell_2/m_2$ $\sim$ 0.5 we obtain for mesogranulation is consistent with this. The different sensitivities of the two components are consistent with their different depths. At the photosphere, radiation takes over from convection and the Mach number is set mainly by the local density, depending on $T_{\rm eff}$ only to a lesser degree \citep{Tremblay2013}. In the subsurface layers, convection carries essentially the whole flux $\sigma T_{\rm eff}^4$, and the horizontal scale of the flow follows the local density scale height, which increases with depth \citep{Beeck2013}. Granulation arises in the former regime and mesogranulation, if it originates in deeper layers, in the latter. Our $m_i-\ell_i$ decomposition is a consistency argument and not a unique measurement, since it does not consider stellar mass, metallicity or the full radiative-transfer response of $\mathcal{M}_i, \eta_i$. Comparing surface Mach numbers and granule sizes from 3D convection simulations across the $T_{\rm eff}–\log g$ plane with these exponents would test it directly. Either way, it shows that the two components do not respond to $T_{\rm eff}$ in the same way, as a single self-similar convective process would require.

\subsection{A shared gravity dependence}
Surface gravity is the one parameter on which granulation and mesogranulation agree: their amplitude and frequency exponents ($\beta_a$ = -1.169 and -1.122; $\beta_b$ = 0.905 and 0.958) are nearly identical and close to the expected values of $\mp$1 if both length scales follow the pressure scale height. The small departures from $\mp$1 have the same direction for both components: the amplitudes are slightly steeper than g$^{-1}$ and the frequencies slightly shallower than g$^{1}$. Both follow if the Mach number decreases weakly with g, since a lower Mach number reduces both the temperature fluctuation, and hence the amplitude, and the flow speed, and hence the frequency. This agrees with 3D simulations, in which higher gravity gives a higher photospheric density and the Mach number is anticorrelated with density, while the granule size scales approximately as g$^{-1}$ at fixed $T_{\rm eff}$, as does $H_p$ \citep{Tremblay2013}. Such fits to simulations are calibrated over a different range of stellar parameters (mainly dwarfs and white dwarfs), and provide no universal gravity exponent for the Mach number, so we use them only qualitatively. Scatter in the spectroscopic $\log g$ also flattens the fitted exponents, which may account for part of the shallower frequency exponents, so we do not interpret the departures quantitatively.

\subsection{The role of metallicity} \label{sec: metallicity discussion}

Metallicity plays a substantial role in both intrinsic amplitudes: $\gamma = 0.350 \pm 0.011$ ($a_{2,\rm int}$) and $\gamma = 0.261 \pm 0.010$ ($a_{3,\rm int}$), both decisively preferred by the Bayesian model comparison. The values are in qualitative agreement with \citet{Corsaro2017}, who found much stronger exponents ($\sim$ 0.89 for $a_{2,\rm obs}$ using a much narrower ($\sim$0.4~dex), metal-rich, cluster-based sample.  Our sample spans a wider, more heterogeneous range ($-2.0$ to $+0.5$~dex). The characteristic frequencies show a weak metallicity dependence: a small, positive exponent for $b_2$ ($\gamma = 0.034 \pm 0.008$), and no statistically justified dependence for $b_3$.

This conclusion is independently supported by both of our validation methods. The GP comparison (Section \ref{sec:gp_validation}) now favors retaining metallicity as a relevant predictor for both amplitudes, consistent with the Bayesian result. Symbolic regression, searching with no assumed functional form, independently recovers substantial, positive metallicity exponents for both amplitudes ($\alpha_{\rm [Fe/H]} = 0.273$ for $a_{2,\rm int}$, $0.269$ for $a_{3,\rm int}$), matching the Bayesian values in both sign and magnitude.

Metallicity therefore discriminates between the amplitudes and the frequencies, rather than between granulation and mesogranulation: it plays a substantial role wherever the disk-integrated brightness contrast is the observable (both amplitudes), and a weak or negligible role wherever the local convective turnover rate is the observable (both frequencies). We discuss the physical interpretation of this pattern, and its relation to the mechanisms proposed in the prior literature, in Appendix~\ref{app:metallicity}.

\section{Conclusions}
We tested the self-similarity of granulation and mesogranulation in a large sample of Kepler RGB and RC stars, with stellar parameters obtained from the APOGEE survey. We find that both convective components depend on surface gravity in a statistically indistinguishable manner ($\beta \simeq -1.1$ to $-1.2$ for the radius-corrected amplitudes, $\beta \simeq 0.9$--$1.0$ for the frequencies), consistent with a simple disk-cancellation and convective-turnover framework in which both components share the same envelope geometry, but their dependence on $T_{\rm eff}$ differs sharply: mesogranulation's amplitude is strongly sensitive to $T_{\rm eff}$ ($\alpha = 1.295$), whereas granulation's shows no statistically justified dependence, while both frequencies show a comparatively weak and similar $T_{\rm eff}$ dependence. We show that this pattern is internally consistent with a picture in which granulation forms in the thin, near-surface, radiatively efficient layer where the Mach number is set mainly by the local density, while mesogranulation originates from deeper, more nearly adiabatic convection-dominated motions whose signal reaches the photosphere through overshoot, rather than representing the same convective process reorganized at a larger spatial scale. Gaussian Process and symbolic regression, neither of which assumes the power-law form, recover the same dependencies. Metallicity plays a substantial role in both amplitudes ($\gamma = 0.350$ and $0.261$) but not in the frequencies, as expected if the Mach number and cell size both rise with [Fe/H], which reinforces them in the amplitude and largely cancels them in the frequency. Together, these results indicate that the self-similarity assumption implicit in the two-component background models widely used in red giant asteroseismology does not hold, with direct implications for the physical interpretation of convective background models and their downstream use in Galactic archaeology.

\begin{acknowledgments}
We acknowledge support from the Department of Atomic Energy, Government of India, under Project Identification No. RTI 4002. This research was supported in part by generous donations from the Murty Trust and Premji Invest, both aimed at enabling advances in astrophysics through the use of machine learning. Murty Trust, an initiative of the Murty Foundation, is a not-for-profit organisation dedicated to the preservation and celebration of culture, science, and knowledge systems born out of India. The Murty Trust is headed by Mrs. Sudha Murty and Mr. Rohan Murty. AM acknowledges support from the Spanish Ministry of Science and Innovation with the the proyecto plan nacional \textit{PLAtoSOnG} (grant no. PID2023-146453NB-I00).
\end{acknowledgments}

\appendix
\twocolumngrid

\begin{deluxetable*}{lccc}
\tablecaption{Consolidated Bayesian model comparison for the granulation
and mesogranulation parameters.
\label{tab:bayesian_model_comparison}}
\tablewidth{\textwidth}
\tablehead{\colhead{Model} &\colhead{$N_{\rm pred}$} & \colhead{$\Delta\ln Z$} &\colhead{$O_{i,\mathrm{best}}$}}
\startdata
\multicolumn{4}{c}{$a_{2, \rm int}$ (mesogranulation amplitude)}\\
\tableline
$T_{\rm eff}$& 1 & $-9932.438$  & $8.31\times10^{-4315}$ \\
$\log g$& 1 & $-468.551$& $1.08\times10^{-204}$ \\
$[\mathrm{Fe/H}]$& 1 & $-11700.004$  & $1.89\times10^{-5082}$ \\
$T_{\rm eff},\log g$ & 2 & $-465.906$ & $1.52\times10^{-203}$ \\
$T_{\rm eff},[\mathrm{Fe/H}]$& 2 & $-9297.334$  & $5.52\times10^{-5039}$ \\
$\log g,[\mathrm{Fe/H}]$& 2 & $-98.910$   & $3.69\times10^{-44}$ \\
$\boldsymbol{T_{\rm eff},\log g,[\mathrm{Fe/H}]}$& \textbf{3} & $\mathbf{0.000}$&$\mathbf{1.000}$ \\
\tableline
\multicolumn{4}{c}{$a_{3, \rm int}$ (granulation amplitude)}\\
\tableline
$T_{\rm eff}$& 1 & $-8933.218$  & $ 2.25\times10^{-3880}$ \\
$\log g$& 1 & $-313.592$   & $6.44\times10^{-137}$ \\
$[\mathrm{Fe/H}]$ & 1 & $ -10890.082$  & $3.14\times10^{-4730}$ \\
$T_{\rm eff},\log g$ & 2 & $-214.036$ & $1.11\times10^{-93}$ \\
$T_{\rm eff},[\mathrm{Fe/H}]$& 2 & $-8220.855$  &$5.35\times10^{-3571}$ \\
$\boldsymbol{\log g,[\mathrm{Fe/H}]}$& \textbf{2} & $\mathbf{0.000}$& $\mathbf{1.000}$ \\
$T_{\rm eff},\log g,[\mathrm{Fe/H}]$& 3 & $-3.012$ & $0.15$ \\
\tableline
\multicolumn{4}{c}{$b_2$ (mesogranulation frequency)}\\
\tableline
$T_{\rm eff}$& 1 & $-10577.222$ & $9.59\times10^{-4594}$ \\
$\log g$& 1 & $-20.133$& $5.99\times10^{-10}$ \\
$[\mathrm{Fe/H}]$& 1 & $-12085.682$ & $1.26\times10^{-5249}$ \\
$T_{\rm eff},\log g$& 2 & $-2.765$& $2.09\times10^{-2}$ \\
$T_{\rm eff},[\mathrm{Fe/H}]$& 2 & $-9393.617$  & $9.60\times10^{-4081}$ \\
$\log g,[\mathrm{Fe/H}]$ & 2 & $-3.041$& $1.59\times10^{-2}$ \\
$\boldsymbol{T_{\rm eff},\log g,[\mathrm{Fe/H}]}$& \textbf{3} & $\mathbf{0.000}$ & $\mathbf{1.000}$ \\
\tableline
\multicolumn{4}{c}{$b_3$ (granulation frequency)}\\
\tableline
$T_{\rm eff}$ & 1 & $-10807.035$ & $1.83\times10^{-4694}$ \\
$\log g$& 1 & $-12.921$& $2.45\times10^{-6}$ \\
$[\mathrm{Fe/H}]$& 1 & $-12360.405$ & $1.19\times10^{-5368}$ \\
$\boldsymbol{T_{\rm eff},\log g}$&\textbf{2} & \bf{0.000} & \bf{1.000} \\
$T_{\rm eff},[\mathrm{Fe/H}]$& 2 & $-9693.587$  & $1.62\times10^{-4210}$ \\
$\log g,[\mathrm{Fe/H}]$& 2 & $-16.823$& $4.94\times10^{-8}$ \\
$T_{\rm eff},\log g,[\mathrm{Fe/H}]$& 3&$-3.745$& $7.12\times10^{-2}$ \\
\enddata
\tablecomments{
$N_{\rm pred}$ is the number of predictors.
For each quantity,
$\Delta\ln Z_i=\ln Z_i-\ln Z_{\rm best}$.
The posterior odds $O_{i,\mathrm{best}}$ are normalized to the
best-supported model and include the adopted combinatorial model prior.
The preferred model for each quantity is shown in bold.
}
\end{deluxetable*}

\section{SAMPLE SELECTION FIGURES} \label{app:sample}
\begin{figure}[!h]
\centering
\includegraphics[width=\columnwidth]{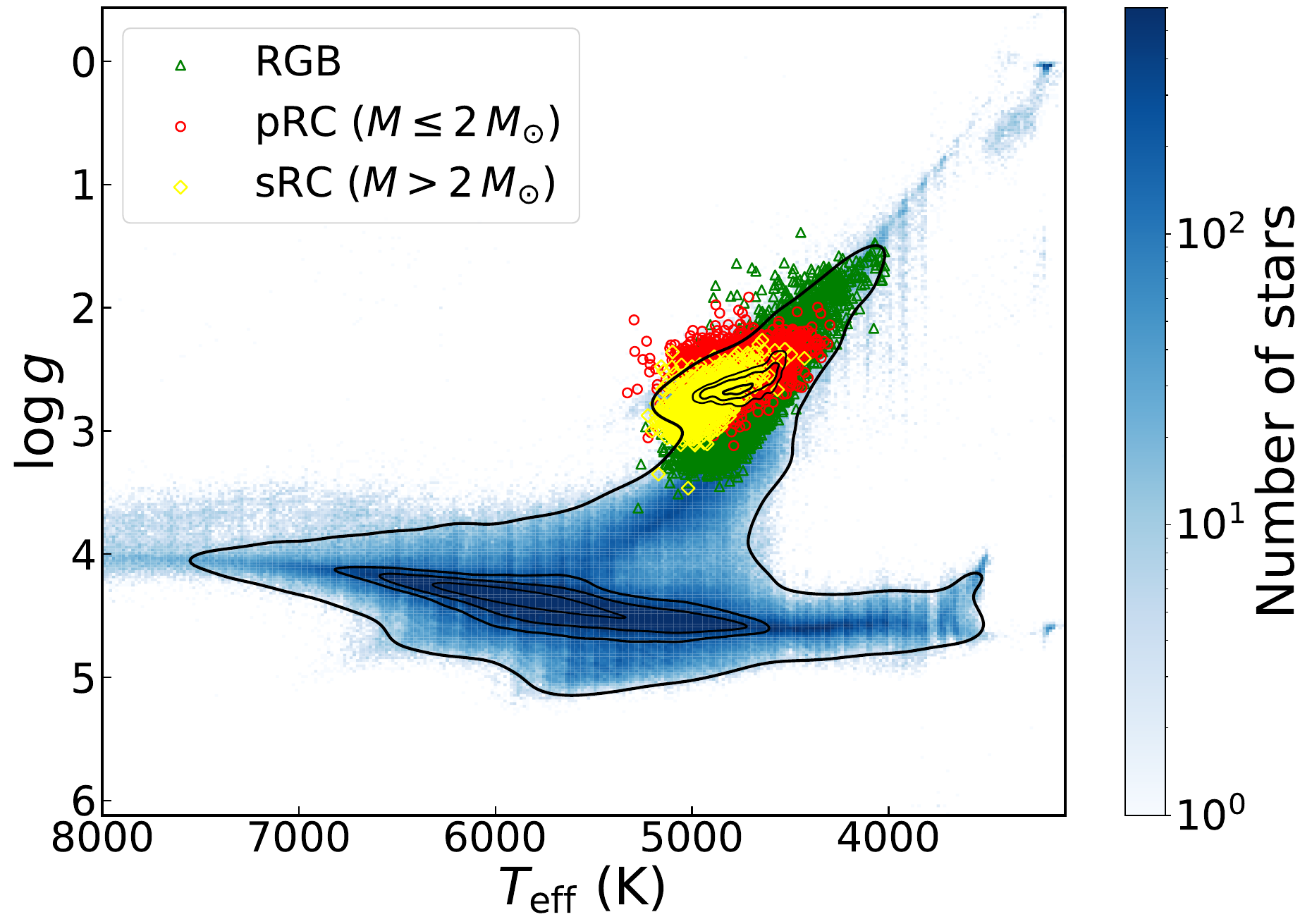}
\caption{Distribution of sample stars in the $T_{\rm eff}$--$\log g$ plane. The blue background shows the density of the entire $Kepler$ sample. Green triangles mark RGB stars, red circles indicate pRC ($M \le 2 M_\odot$) stars, and yellow diamonds denote sRC ($M > 2 M_\odot$) stars.}
\label{fig:1}
\end{figure}

\begin{figure}[!h]
\centering
\includegraphics[width=\columnwidth]{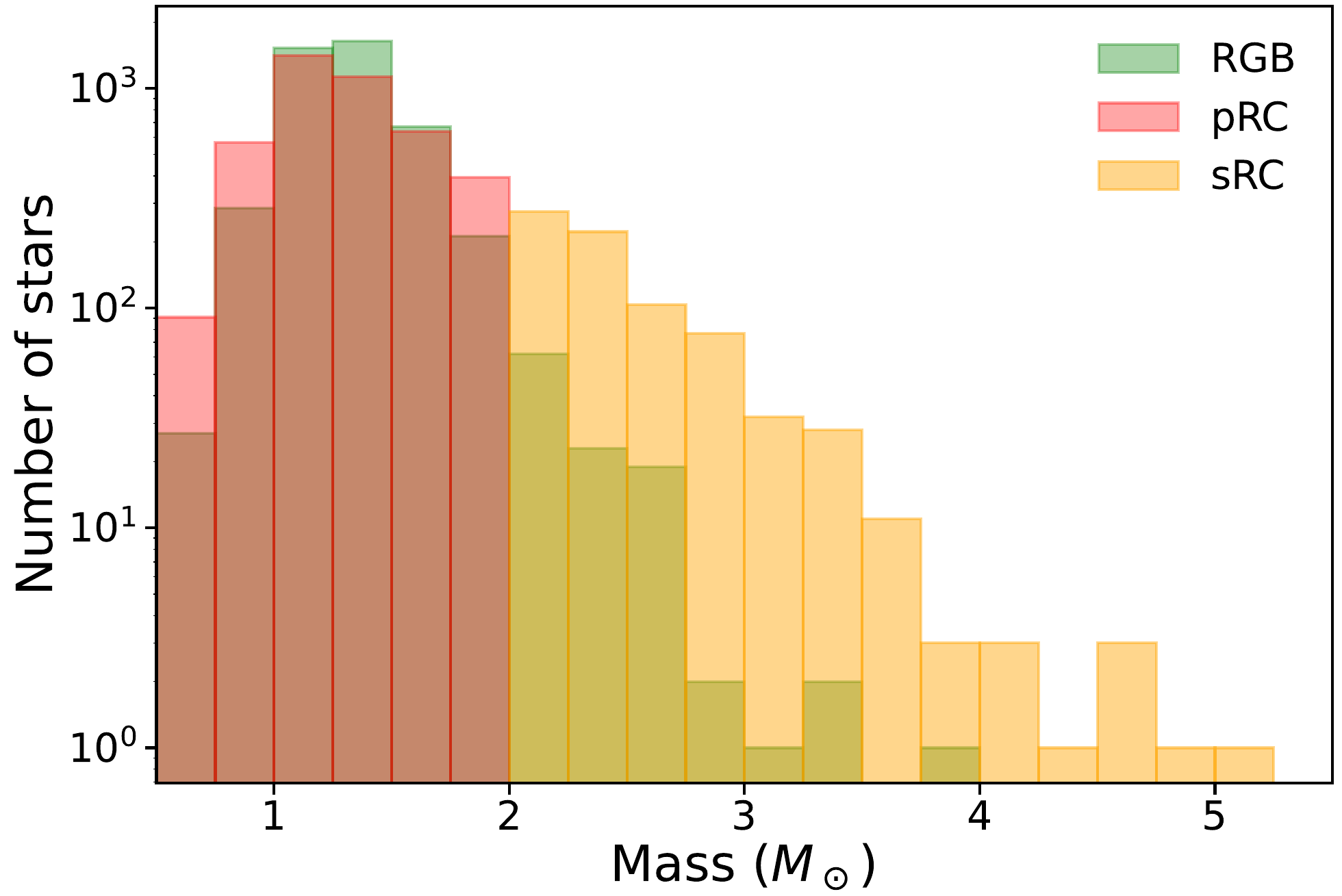}
\caption{Comparison of the asteroseismic stellar mass distributions for RGB, pRC, and sRC stars.}
\label{fig:mass distribution}
\end{figure}
Figure~\ref{fig:1} and Figure~\ref{fig:mass distribution} show the distribution of our final sample in the $T_{\rm eff}$--$\log g$ plane, color-coded by evolutionary phase and overplotted on the density of the full $Kepler$ input catalog, together with the corresponding distribution of asteroseismic stellar mass for each evolutionary sub-population (Section \ref{sec: sample and data}).

\section{Background-fitting validation against KIC 7949599} \label{app:validation}

As a validation of our background-fitting methodology against previously published results, we compare our fit for KIC~7949599 with the parameters reported by \citet{Kallinger2014}, who fit this star's power spectrum as part of their original demonstration of the two-component granulation background model. They report $a_2 = 466\pm14$~ppm, $b_2 = 9.4\pm0.5\,\mu$Hz, $a_3 = 399\pm19$~ppm, and $b_3 = 31.9\pm1.0\,\mu$Hz. Our fit yields $b_2 = 9.64^{+0.44}_{-0.43}\,\mu$Hz and $b_3 = 33.49^{+0.71}_{-0.61}\,\mu$Hz, differing by only $0.36\sigma$ and $1.36\sigma$ respectively. Our amplitudes, $a_2 = 470.11^{+13.28}_{-13.81}$~ppm and $a_3 = 415.67^{+16.59}_{-17.33}$~ppm, differ by only $0.21\sigma$ and $0.65\sigma$. We attribute these small offsets to differences in power spectrum construction (our spectra are built from KEPSEISMIC PSDs \citep{Garcia2011}, while \citet{Kallinger2014} use a distinct light-curve correction pipeline) rather than to the fitting procedure itself.

\section{Bayesian model comparison: full results} \label{app:model_comparison}

Table~\ref{tab:exponents NSMC} adopts the model with the highest posterior odds ratio, $O_{j,{\rm best}}$, for each quantity, computed as the Bayes factor between model $j$ and every other candidate model, weighted by a combinatorial prior that disfavors unnecessarily complex models. Table \ref{tab:bayesian_model_comparison} gives the Bayesian evidences ($ \Delta \ln Z$), and odds ratios relative to the best-supported model, for all seven candidate models and all four quantities. 

\section{Gaussian Process validation: full results} \label{app:gp}
We use a GP whose covariance is a squared-exponential (RBF) kernel with a separate length scale for each predictor, plus a white-noise term,
\begin{equation}
  k(\mathbf{x},\mathbf{x}') = \sigma_f^2
  \exp\!\Big[-\frac{1}{2}\sum_{j}\frac{(x_j-x'_j)^2}{\ell_j^2}\Big]
  + \sigma_n^2\,\delta_{\mathbf{x}\mathbf{x}'},
  \label{eq:gpkernel}
\end{equation}
where $j$ runs over the predictors of the adopted model, each standardized to zero mean and unit variance (computed on the GP training sample). The target, $\ln(X/X_\odot)$, is standardized in the same way. Here
$\sigma_f$ is the signal amplitude, $\sigma_n$ is the amplitude of the white noise, which absorbs scatter beyond the measurement uncertainties, and $\ell_j$ is the length scale of predictor $j$ in units of its standard deviation. A small $\ell_j$ means the quantity varies rapidly along that axis, so the GP correction depends strongly on that predictor; a large $\ell_j$
means it is nearly constant along that axis.

We randomly withheld 20\% of the parent sample before training, giving 1866 test stars representative of the full RGB and RC sample; an additional 1500 stars were selected randomly for GP training. GP training of the full sample was not performed to reduce computational time. The adopted parametric coefficients were obtained from the full-sample nested-sampling analysis, so the comparison slightly favors the parametric relations and the measured GP improvements may be regarded as conservative. For each observable, we fitted GP models for all seven predictor combinations (in Table \ref{tab:bayesian_model_comparison} and selected the combination with the lowest log-space RMSE on the test data. The selected combinations matched those favoured by Bayesian model comparison using UltraNest. We then quantified the improvement of each selected GP model over its corresponding parametric relation as the percentage reduction in log-space RMSE (see Section \ref{sec:gp_results}).

Table~\ref{tab:gp_summary} compares the predictive performance of the GP and the adopted parametric relations, evaluated in $\ln(X/X_\odot)$. Figure~\ref{fig:gp_obspred} shows the residuals of the GP and parametric predictions against the parametric prediction. For \(a_{2,\mathrm{int}}\), the median residual of the parametric relation changes from negative to slightly positive and back to negative across the predicted-amplitude range; a weaker variation is visible for \(a_{3,\mathrm{int}}\). The GP median residuals lie closer to zero in some intervals, suggesting a modest reduction in systematic bias, although the two median curves overlap substantially. These trends alone do not establish how much of the amplitude improvement arises from reduced curvature rather than changes in scatter or large residuals. For \(b_2\) and \(b_3\), both median curves also remain close to zero, with small variations across the predicted-frequency range.

The optimized GP length scales indicate the nonlinear correction is governed mainly by surface gravity: in standardized coordinates, $\ell_j$ in $\ln(g/g_\odot)$ are 3.05, 5.56, 1.67, and 0.82 for $a_{2, \rm int}, a_{3, \rm int}, b_2$ and $b_3$ respectively, substantially shorter than the corresponding $T_{\rm eff}$ length scales (8.53, 11.4, 7.45, 8.92). The metallicity length scales (14.2, 35.3, 48.1, 69.9) are much longer than the range spanned by the
data in standardized units, indicating only weak additional nonlinear metallicity structure compared with $T_{\rm eff}$.

\begin{figure*}[!htb]
\centering
\includegraphics[width=0.9\textwidth, height = 9cm]{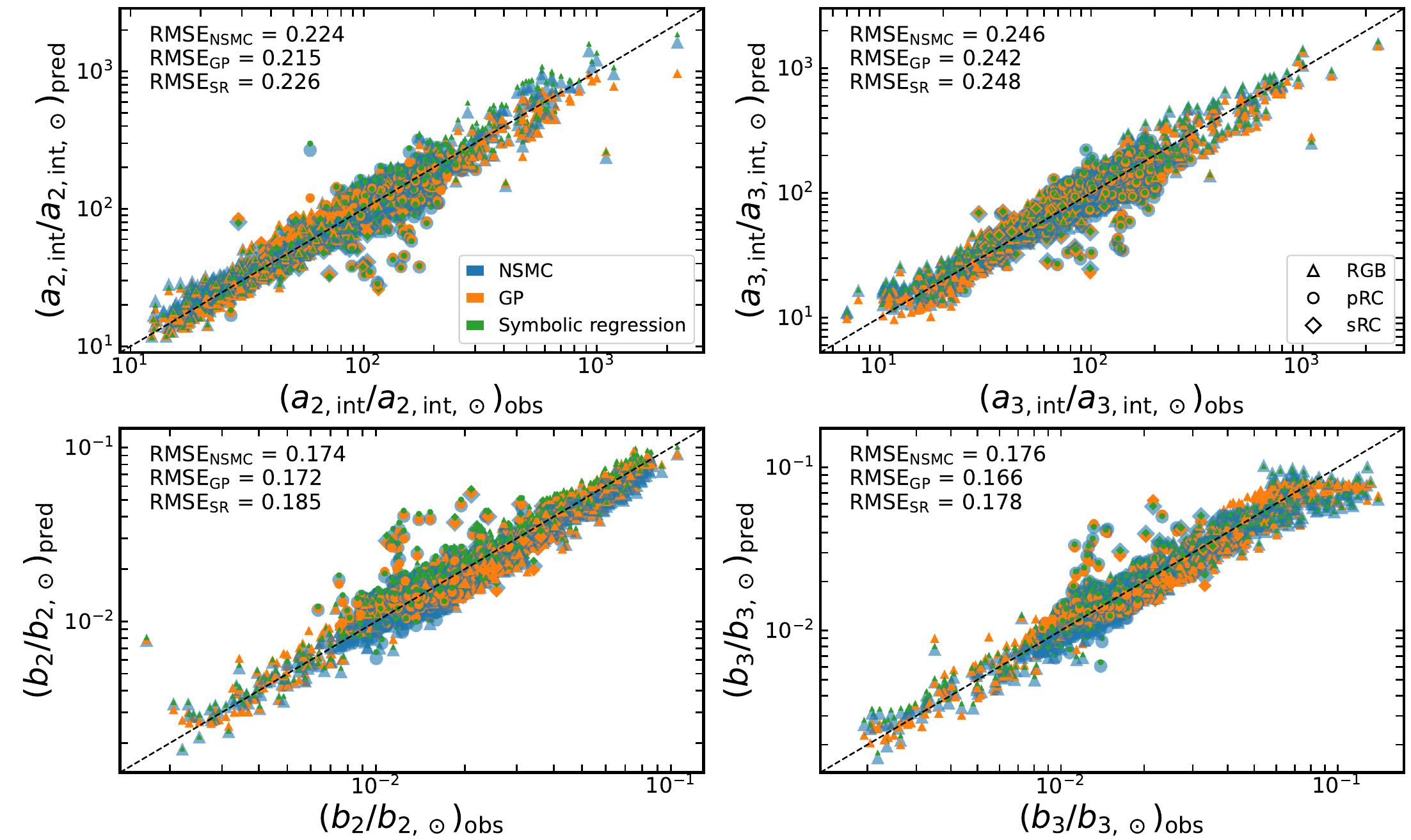}
\caption{Observed versus predicted normalized amplitudes and characteristic frequencies for the GP test sample. UltraNest (Bayesian), GP, and symbolic-regression predictions are shown in blue, orange, and green, respectively; symbols denote RGB, pRC, and sRC stars. RMSE values are given for each panel.}
\label{fig:threeway_comparison}
\end{figure*}

\begin{deluxetable*}{lccccc}
\tabletypesize{\scriptsize}
\tablecaption{Comparison between the GP predictions and adopted parametric scaling relations on the test sample. All quantities in $\ln(X/X_\odot)$. \label{tab:gp_summary}}
\tablewidth{0pt}
\tablehead{
\colhead{Quantity} & \colhead{RMSE$_{\rm par}$} & \colhead{RMSE$_{\rm GP}$} & \colhead{Improvement} & \colhead{$C_{1\sigma}$} & \colhead{$C_{2\sigma}$}
}
\startdata
$a_{2, \rm int}$ & 0.224 & 0.215 & $3.9\%$ & $73.1\%$ & $95.0\%$ \\
$a_{3, \rm int}$ & 0.246 & 0.244 & $0.9\%$ & $69.4\%$ & $94.7\%$ \\
$b_2$ & 0.175 & 0.169 & $3.4\%$ & $74.1\%$ & $95.3\%$ \\
$b_3$ & 0.181 & 0.174 & $3.8\%$ & $75.3\%$ & $95.4\%$ \\
\enddata
\tablecomments{Improvement is $100(\mathrm{RMSE}_{\rm par}-\mathrm{RMSE}_{\rm GP})/\mathrm{RMSE}_{\rm par}$; $C_{1\sigma}, C_{2\sigma}$ are the GP coverage fractions. Standard deviations of the normalized GP pull range from 1.03--1.07.}
\end{deluxetable*}

\section{Symbolic regression: full results} \label{app:sr}
PySR scores each expression by its
prediction loss and complexity, and returns the best equation at each complexity. Each star is weighted by $w_i = 1/\sigma_i^2$, where $\sigma_i$ is its measurement uncertainty in $\ln X$ (the upper and lower errors converted to logarithmic space and averaged), normalized to a mean weight of unity, and the loss is the weighted mean squared error. We repeat the
search with five random seeds (11, 22, 33, 44, 55), which initialize the random number generator and hence the search trajectory, using PySR's deterministic
serial mode so that each seed is reproducible. For each quantity we keep the run with the smallest weighted RMSE of the log-space residuals, $\mathrm{RMSE}_w=\big[\sum_i w_i r_i^2/\sum_i w_i\big]^{1/2}$ with $r_i=y_i-\hat y_i$, evaluated on the full sample. The equation selected from
that run is the one chosen by PySR's default criterion (the highest score, a measure of loss reduction per unit of added complexity, among the equations with a loss close to the best).

Table~\ref{tab:sr_summary} lists the selected equation's recovered exponents and $\sigma_{\rm int}$ for each quantity. 
For all four quantities, the symbolic regression loss drops steeply between the null (constant) model and the first non-trivial equation. The selected intrinsic amplitude equations includes [Fe/H] but not the frequencies, despite this parameter being retained in the Bayesian best-fit model for $b_2$; equations including [Fe/H] for $b_2$ appear at higher complexity but yield only marginal loss reduction, consistent with the small metallicity exponent found in the Bayesian analysis. Across all four quantities, the intrinsic scatter recovered from symbolic regression is systematically larger than the corresponding Bayesian value (e.g., 0.220 vs. 0.190 for $a_2$), consistent with the symbolic regression $\sigma_{\rm int, SR}$ being estimated independently via maximum-likelihood optimization rather than a full posterior framework. PySR selects expressions according to prediction loss and complexity, whereas the Bayesian analysis fits a prescribed scaling relation while accounting for measurement uncertainties and intrinsic scatter. Despite these methodological differences, the broadly similar coefficients support the robustness of our inferred scaling dependencies.
Figure~\ref{fig:threeway_comparison} shows a direct visual comparison of all three modeling approaches, the Bayesian scaling relations, the GP, and symbolic regression,  plotted against the observed value for each quantity.

\begin{deluxetable*}{lcccccccccc}
\tablecaption{Symbolic regression results compared to the Bayesian scaling relations \label{tab:sr_summary}}
\tablewidth{\textwidth}
\tablehead{
\colhead{Parameter} & \colhead{$N$} & \colhead{Best Seed} & \colhead{Complexity} &\colhead{Loss} & \colhead{RMSE$_{\log}$} & \colhead{$C_{\rm SR}$}& \colhead{$\alpha_{\rm SR}$} & \colhead{$\beta_{\rm SR}$} & \colhead{$\gamma_{\rm SR}$} & \colhead{$\sigma_{\rm int,SR}$}
}
\startdata
$a_{2, \rm int}$ (meso) & 9328 & 55 & 13 & 0.036& 0.228 & 0.761& 1.000 & $-1.198$ & 0.273 & 0.220 \\
$a_{3, \rm int}$ (gran) & 9328 & 33 & 9 & 0.044& 0.246 & 0.764 & \nodata & $-1.128$ & 0.269 & 0.236 \\
$b_2$ (meso) & 9328 & 44 & 9 & 0.032& 0.168 & 0.955 &$-0.202$ & 0.924 & \nodata & 0.159 \\
$b_3$ (gran) & 9328 & 22 & 9 & 0.023& 0.172 & 0.979 & $-0.299$ & 0.937 & \nodata & 0.161 \\
\enddata
\tablecomments{$N$ is the number of stars used; ``Best Seed'' is the random-number seed (one of 11, 22, 33, 44, 55) of the run with the lowest inverse-variance-weighted RMSE of the log-space residuals;; Complexity is the total number of operators, variables, and constants in the selected equation. Loss is the mean-squared error of the selected equation on the training data, as reported by PySR. $C_{\rm SR}$ is the normalization, defined as in Equation \ref{eq:form}; $\alpha_{\rm SR}, \beta_{\rm SR}$, $\gamma_{\rm SR}$ are the $T_{\rm eff}$, $\log g$ and [Fe/H] exponents respectively; $\sigma_{\rm int,SR}$ is the corresponding intrinsic scatter. \nodata indicates the parameter was not included in the best-supported model.Compare to Table~\ref{tab:exponents NSMC}.}
\end{deluxetable*}

\section{EXTENDED DISCUSSION OF THE METALLICITY DEPENDENCE}
\label{app:metallicity}

Section \ref{sec: metallicity discussion} shows that metallicity discriminates between the amplitudes and the frequencies, not between granulation and mesogranulation. Here we expand on the supporting evidence, a qualitative physical interpretation and give a comparison with previous works.

Both amplitudes retain a positive metallicity exponent, and symbolic regression independently selects a [Fe/H] term in both. Withholding [Fe/H] raises the RMSE of the held-out test stars by $\approx$3\% for both the parametric relations and the GP (slightly less for the $a_{2,\rm int}$ GP, whose stronger dependence on $T_{\rm eff}$ can partly stand in for metallicity). For the frequencies, only $b_2$ retains a small positive term in the Bayesian analysis ($\gamma_{b_2} = 0.034 \pm 0.008$), and neither the GP nor symbolic regression supports it; none is retained for $b_3$.

The pattern follows from how metallicity enters the convective properties of Section \ref{sec: Teff discussion}. In 3D simulations at fixed $T_{\rm eff}$ and $\log g$, higher metallicity raises the opacity and lowers the photospheric density, which raises the Mach number, and the granule size and intensity contrast rise with it \citep{Tremblay2013}. Both $\mathcal{M}_i$ and $\eta_i$ therefore increase with [Fe/H]. The amplitude, $a_{i,\rm int} \propto \mathcal{M}_i^2\eta_i$, is a product, so the two increases reinforce each other and metallicity matters for both components. The frequency, $b_i \propto \mathcal{M}_i/\eta_i$, is a ratio, so the two increases largely cancel, leaving it much less sensitive. This is why the frequencies show at most a weak trend: the one for $b_2$ is not supported by the independent methods and we treat it as marginal, and none is needed for $b_3$, whose frequency lies closest to the baseline expected for constant $\mathcal{M}_i$ and $\eta_i$.

The same simulations suggest why the amplitude exponent is somewhat larger for mesogranulation than for granulation. At a given Mach number, the intensity contrast is higher at low metallicity, because convection is more efficient, which partly offsets the rise of the contrast with [Fe/H] \citep{Tremblay2013}. This offset arises from radiative cooling at the surface, so it should weigh more on granulation, which forms there, than on the deeper flows we associate with mesogranulation, and would lower the net exponent of granulation relative to mesogranulation. We have not tested this explanation, and the difference between the two exponents is modest.

Comparisons with \citet{Corsaro2017} warrant care: their analysis uses 36 red giants in three open clusters, with metallicity calibrated from cluster means over a narrow, metal-rich range ($-0.09$ to $+0.32$~dex, $\sim$0.4~dex total), reducing age/evolutionary-state scatter at the cost of metallicity coverage. Our field sample spans $-2.0$ to $+0.5$~dex, at the cost of greater population heterogeneity. Their metallicity coefficient (u=0.89 $\pm$ 0.08) is larger than ours. Another work by \citet{Rodriguez2022} shows a value u=0.211 $\pm$ 0.033 (from 3D simulations with a single convective scale) in range with our reported values. But these relations are at fixed $\nu_{\rm max}$ and are not strictly comparable to ours at fixed $T_{\rm eff}$ and $\log g$ for radius-corrected amplitudes of two components. Our small $\gamma_{b_2}$ is opposite in sign to the longer timescales for metal-rich giants reported by \citet{Corsaro2017}, which we regard as another reason to treat it as a weak trend; 3D models likewise show only a weak metallicity dependence of the autocorrelation time \citep{Rodriguez2022}.

{We caution that because both components are fit simultaneously from overlapping regions of the same power spectrum, a metallicity-dependent redistribution of power between them cannot be entirely excluded, which also limits the interpretation of the difference between the two amplitude exponents. Metallicity is correlated with age and mass in field red giants, and also with $T_{\rm eff}$ at fixed $\log g$, so the [Fe/H] term may partly absorb residual dependences not included in the model; and since all methods use the same APOGEE abundances, their agreement does not test systematics in the abundances themselves. Stronger effects of low metallicity on the upper atmosphere, such as enhanced reversed granulation, have been reported at [Fe/H]~$=-3$ and $T_{\rm eff} \gtrsim 5500$~K \citep{Magic2013}, beyond the range of most of our sample. The absence of a metallicity term for $b_3$ should be read as ``not statistically justified,'' not as establishing strict independence from metallicity.}

\bibliography{references}{}
\bibliographystyle{aasjournalv7}
\end{document}